\documentclass[11pt, oneside]{article}   	
\usepackage[margin=2cm]{geometry}                		
\usepackage[parfill]{parskip}    		
\usepackage{graphicx}				
\usepackage[utf8]{inputenc}
\usepackage[mathlines]{lineno}
\usepackage[english]{babel} 
\usepackage{amsmath,amssymb,amsthm,bm} 
\usepackage[margin=2cm]{geometry}
\usepackage[color=yellow]{todonotes}
\usepackage{url}	
\usepackage{esint}
\usepackage[colorlinks]{hyperref}
\usepackage{enumerate}
\usepackage{outlines}[enumerate]
\usepackage{framed,comment}
\usepackage{upgreek}
\usepackage{pgfplots}
\usepackage{float}
\usepackage{tikz,tikz-cd}
\usepackage{rotating}
\usepackage{graphicx}
\usepackage{caption}
\usepackage{cancel}
\usepackage{subcaption}
\usepackage{appendix}
\usepackage{tikz-cd}
\usepackage [english]{babel}
\usepackage [autostyle, english = american]{csquotes}
\usepackage{pgf, pgffor}
\extrafloats{100}
\numberwithin{equation}{section}

\newtheorem{theorem}{Theorem}[section]

\newtheorem{proposition}{Proposition}[section]

\newcommand{\mb}[1]{\mathbf{#1}}
\newcommand{\mbb}[1]{\mathbb{#1}}
\newcommand{\mc}[1]{\mathcal{#1}}
\newcommand{\bs}[1]{\boldsymbol{#1}}

\newcommand{\ad}{\operatorname{ad}}
\newcommand{\di}{\operatorname{div}}
\newcommand{\dd}{\mathrm{~d}}
\newcommand{\dn}{\mathrm{d}}

\def\p{{\partial}}

\def\ba{{\mathbf{a}}}
\def\bb{{\mathbf{b}}}

\def\bk{{\mathbf{k}}}

\MakeOuterQuote{"}

\title{Variational Principles on Geometric Rough Paths \\and the L\'{e}vy Area Correction}
\author{Theo Diamantakis, Darryl D. Holm, Grigorios A. Pavliotis \\ 
Department of Mathematics, Imperial College London, London SW7 2AZ} 
\date{}

\begin{document}
\maketitle

\begin{abstract}
In this paper, we describe two effects of the L\'evy area correction on the invariant measure of stochastic rigid body dynamics on geometric rough paths. From the viewpoint of dynamics, the L\'evy area correction introduces an additional deterministic torque into the rigid body motion equation on geometric rough paths. When the rigid body dynamics is driven by coloured noise, and damped by double-bracket dissipation,  our theoretical and numerical results show that the additional deterministic torque due to the L\'evy area correction shifts the centre of the probability distribution function by shifting the Hamiltonian function in the exponent of the Gibbsian invariant measure.  
\end{abstract}

\newpage
\section{Introduction}\label{sec:Intro}

Ideal fluid dynamics on geometric rough paths (GRP) has recently been studied in two companion papers \cite{Leahy2020,Crisan2022}. The first of these papers \cite{Leahy2020} derives ideal fluid equations on GRP by using constrained variational principles that respect the geometric structure of ideal fluid dynamics. This geometric structure arises from Lie symmetry reduction from Lagrangian to Eulerian fluid variables in the variational principle. The Lie symmetry reduction of these variational principles yields a GRP version of the Euler-Poincar\'e formulation of deterministic 3D ideal fluid dynamics, including its the Kelvin-Noether circulation theorem and its Lie–Poisson Hamiltonian formulation \cite{holm1998euler}. The GRP formulation of ideal fluid dynamics in \cite{Leahy2020} enhances its stochastic counterpart derived in \cite{Holm2015} and extended to semimartingales in \cite{street2021semi}. 

The second of these companion papers \cite{Crisan2022} continues by proving the local well-posedness of Euler’s ideal fluid equations on GRP in 3D and by establishing a Beale-Kato-Majda (BKM) blow-up criterion in terms of the time-integrated $L^\infty$ norm of the vorticity. In particular, the second paper \cite{Crisan2022} shows that Euler’s equations for 3D ideal fluid dynamics on GRP possess the same BKM blow-up criterion as in the deterministic case.

In deriving variational ideal fluid dynamics on GRP and establishing its solution properties, the two companion papers \cite{Leahy2020,Crisan2022} extended the corresponding results for stochastic partial differential equations (SPDEs) that had been proven earlier for 3D ideal fluid SPDE in \cite{Holm2015} and \cite{crisan2019solution} respectively.
These earlier results for fluid dynamical SPDE with stochastic advection by Lie transport (SALT) have led to new algorithms for estimating uncertainty in coarse-grained fluid flows and the further development of data calibration and assimilation methods that effectively reduce the uncertainty of stochastic coarse-grained fluid models \cite{cotter2018modelling,cotter2019numerically}. 

The 3D Euler fluid equations with SALT can be written using the Leray operator $\mbb{P}$ to project onto the divergence-free part  of its operand as
\begin{align}
\begin{split}
\dn \mb{v}(\mb{x},t) &=  \mbb{P} \bigg(\Big(\mb{v}(\mb{x},t) \dn t 
+ \sum_{i=1}^N \boldsymbol{\xi}_i(\mb{x}) \circ \dn W^i_t\Big) \times {\rm curl}\mb{v}(\mb{x},t)\bigg)
\\&= 
\mbb{P} \bigg( \frac{\delta \dn H}{\delta \mb{v}} \times \frac{\delta C}{\delta \mb{v}} \bigg)
,\\ \hbox{with stochastic Hamiltonian } \dn H(\mb{v}) 
&=   \int_{\mbb{R}^3}\bigg( \frac12\mbb{P}\mb{v}\cdot \mb{v} \ \dn t 
+ \mbb{P}\mb{v}(\mb{x},t) \cdot \sum_{i=1}^N \boldsymbol{\xi}_i(\mb{x}) \circ \dn W^i_t \bigg) d^3x 
\\ \hbox{and conserved helicity } C(\mb{v}) &=  \frac12\int_{\mbb{R}^3} \mb{v}(\mb{x},t) \cdot {\rm curl}\,\mb{v}(\mb{x},t)\, d^3x 
\,,
\end{split}
\label{eqn: SALTEuler}
\end{align}
where $\delta/\delta \mb{v}$ represents variational derivative with respect to the divergence-free fluid velocity, $\mb{v}$, and the $\boldsymbol{\xi}_i(\mb{x})$, $i=1,\dots,N,$ are prescribed vector-field functions of spatial position $\mb{x}$ independent of time $t$ that are to be determined from data calibration in a given data set, according to the SALT algorithm of  \cite{cotter2018modelling,cotter2019numerically}.

In comparison, the stochastic rigid body equations which are treated in \cite{Arnaudon2018} are given by 
\begin{align}
\begin{split}
    {\rm d}\boldsymbol\Pi &= -\,\Big(\mbb{I}^{-1}\bs{\Pi}\, \dn t + \sum_{i=1}^3 \boldsymbol{\sigma}_i \circ \dn W^i_t\Big) \times \boldsymbol\Pi 
\\&= 
- \, \frac{\p \dn H}{\p \bs{\Pi}} \times \frac{\p C}{\p \bs{\Pi}} 
\,,\\ \hbox{with stochastic Hamiltonian } \dn H(\bs{\Pi}) &=  \bigg( \frac12 \bs{\Pi}\cdot \mbb{I}^{-1}\bs{\Pi}\ \dn t 
+ \bs{\Pi} \cdot \sum_{i=1}^3 \boldsymbol{\sigma}_i(\mb{x}) \circ \dn W^i_t \bigg) 
\\ \hbox{and conserved quantity } C(\bs{\Pi}) &=  \frac12 |\bs{\Pi}|^2 
\,.
    \end{split}
    \label{Sto-RB-SALT}
\end{align}
The rigid body equations in \eqref{Sto-RB-SALT} are very similar in form to the Euler fluid equations in \eqref{eqn: SALTEuler}. 
In equation \eqref{Sto-RB-SALT}, $\boldsymbol\Pi\in\mbb{R}^3$ denotes the body angular momentum and $\boldsymbol\Omega = \mbb{I}^{-1}\boldsymbol\Pi\in\mbb{R}^3$ is the body angular velocity. In \eqref{Sto-RB-SALT}, the body angular velocity has been augmented by a sum of Stratonovich noises with constant amplitudes $\boldsymbol{\sigma}_i\in\mbb{R}^3$, $i=1,2,3$. The real constant $3\times3$ diagonal matrix $\mbb{I}$ denotes the principle moments of inertia in the body. The stochastic rigid body equations in \eqref{Sto-RB-SALT} conserve the squared magnitude of the angular momentum $|\boldsymbol\Pi|^2=\boldsymbol\Pi\cdot\boldsymbol\Pi$. 
Following the theory introduced in \cite{GayBalmaz2014}, a selective dissipation of the rigid body energy $\tfrac12\boldsymbol\Pi\cdot\boldsymbol\Omega:=\tfrac12\boldsymbol\Pi\cdot\mbb{I}^{-1}\boldsymbol\Pi$ at constant angular momentum magnitude $|\boldsymbol\Pi|$ was included in \cite{Arnaudon2018}. This selective energy dissipation was accomplished by adding a double-bracket term to the motion equation in \eqref{Sto-RB-SALT}, as follows
\begin{align}
    {\rm d}\boldsymbol\Pi = -\,\Big(\boldsymbol\Omega\, \dn t + \sum_{i=1}^3 \boldsymbol{\sigma}_i \circ \dn W^i_t\Big) \times \boldsymbol\Pi 
    +\vartheta \big(\boldsymbol\Omega\times \boldsymbol\Pi\big) \times \boldsymbol\Pi \, \dn t
    \,,
    \label{Sto-RB-stra-dbrkt}
\end{align}
in which the positive real constant parameter $\vartheta>0$ determines the rate of energy dissipation. The double-bracket dissipation of energy drives the probability distribution of the stochastic rigid body flow toward an equilibrium near the minimum energy level consistent with remaining on the angular momentum sphere at the constant radius $|\boldsymbol\Pi(t)|=|\boldsymbol\Pi(0)|$ determined by the initial conditions.

The SALT rigid body equations in \eqref{Sto-RB-SALT} and the SALT Euler fluid equations in \eqref{eqn: SALTEuler} are remarkably similar in form. Moreover, both of them admit selective dissipation of energy while preserving another conservation law, as discussed in \cite{GayBalmaz2014}. The selective energy dissipation property guarantees that the invariant measures for either of them cannot have zero energy, provided the conserved quantity $C$ does not vanish initially.

When planning to investigate the solution behaviour of GRP dynamics for fluid dynamics, it stands to reason to take advantage of the structural similarity between the equations for SALT Euler fluid dynamics in \eqref{eqn: SALTEuler} and rigid body motion equation \eqref{Sto-RB-SALT}. Furthermore, it makes sense to introduce GRP dynamics into the simpler rigid body problem first, because one can take advantage of what is already known about stochastic transport in rigid body dynamics. Indeed, investigation of the \emph{stochastic} transport dynamics of rigid body motion in \cite{Arnaudon2018} with double-bracket selective energy dissipation as in \eqref{Sto-RB-stra-dbrkt} has already led  to interesting properties for its invariant measure. Namely, the invariant measure for stochastic rigid  body dynamics with double-bracket energy dissipation in \eqref{Sto-RB-stra-dbrkt} {is Gibbsian}. The question then naturally arises as to whether a similar result holds for the invariant measure of the rigid body dynamics on GRP with selective energy dissipation. Answering {this} question {in the affirmative} was the original aim of the present paper and it remains a motivational example. In following the original direction, though, the present paper has also led to the more general result in Theorem \ref{thm: Gibbs}. We believe that this result opens new interesting avenues of research about the entire class of coadjoint dynamics with selective decay on GRP.

The storyline of the present paper focuses on rigid body dynamics on GRP derived from a variational principle that is invariant under $SO(3)$, the special orthogonal Lie group of rotations in three dimensions. In the classical dynamical system of rigid body motion, invariance under $SO(3)$ rotations of the initial reference configuration carries over to $SO(3)$ invariance of the Lagrangian in Hamilton's variational principle for the motion. As a result of reduction of Hamilton's principle by $SO(3)$ symmetry, the body angular momentum takes values in the dual space $\mathfrak{so}(3)^*$ of the Lie algebra $\mathfrak{so}(3)$ of angular rotation velocities. Since both $\mathfrak{so}(3)$ and $\mathfrak{so}(3)^*$ may be represented by $3\times 3$ antisymmetric matrices under Frobenius pairing, they are both isomorphic to $\mathbb{R}^3$ under Euclidean pairing.

In the dynamics of the rigid body in $\mathbb{R}^3$ under $SO(3)$ rotations whose time dependence follows GRP, the dynamical
rotation paths in $SO(3)\times\mathbb{R}^3\to\mathbb{R}^3$ preserve the magnitude $|\boldsymbol{\Pi}|$ of the body angular momentum vector $\boldsymbol{\Pi}$ in $\mathfrak{so}(3)^*\simeq \mathbb{R}^3$, and the vector cross product $\times$ plays the role of the Lie bracket for the Lie algebra $\mathfrak{so}(3)\simeq \mathbb{R}^3$. As mentioned earlier, the rigid body problem has already been studied as an example in the context of stochastic rotations \cite{Arnaudon2018}. In what follows, though, our discussions will include the additional effects of the L\'evy area drift terms induced in rigid body dynamics on geometric rough paths on the Lie group manifold of $SO(3)$.

{\color{black}The theory in \cite{Leahy2020} enables one to write rigid body equations on GRP as

\begin{align}
    {\rm d}\boldsymbol\Pi = -\,\big(\boldsymbol\Omega\, \dn t + \sum_{i=1}^3 \boldsymbol{\sigma}_i \dn \mb{Z}^i_t\big) \times \boldsymbol\Pi \, ,
    \label{GRP-RB}
\end{align}
after making suitable choices of variational principle and $SO(3)$ invariant Lagrangian. The replacement of the Stratonovich differential $\circ \dn W_t$ by the GRP $\dn \mb{Z}_t$ as the driver in \eqref{GRP-RB} leaves a degree of freedom still unspecified, though, since rough paths are always defined using additional data constructed from iterated integrals. For example, the Brownian-motion driven SDE 
\eqref{Sto-RB-SALT} has the driving path $W_t$ specified, but the area between a Brownian curve and its chord for two times $t,s$ (i.e., the L\'evy stochastic area) cannot be determined solely from knowledge of the path $W_t$. See \cite{lejay2009yet} [Sec. 4]. The L\'evy stochastic area is  the antisymmetric iterated integral defined by 
\begin{equation}
    \operatorname{Alt}(\mathbb{B})^{ij}_{0,t}:= \int_0^t B_s^i \circ \dn B_s^j - B_s^j \circ \dn B_s^i \,.
\label{def: LevyArea}
\end{equation} 
While the symmetric iterated integrals must be fixed as a consequence of the assumption that the rough path is geometric, the antisymmetric iterated integral defining the L\'evy stochastic area in \eqref{def: LevyArea} remains arbitrary up to the addition of a deterministic antisymmetric matrix $\mathsf{s}$. Throughout this paper we will refer to choosing $\mathsf{s} = 0$ as the \textit{canonical} choice of L\'evy area as in \eqref{def: LevyArea}. However, in what follows, we will be considering the more general case of \eqref{GRP-RB} in which the arbitrary perturbation $\mathsf{s}$ is retained. }

The theory of GRP tells us that a change in the L\'evy area can be interpreted as a different interpretation of the stochastic integral. The interpretion of the stochastic integral in the context of stochastic fluid dynamics has been considered in \cite{Holm2020} , where correction terms between the It\^o and Stratonovich integrals were found to have physical meaning as the presence of a Coriolis force and importantly, the It\^o-Stratonovich term maintained the geometric structure of the stochastic Euler equations. The aim of this paper is to provide a similar overview of how the L\'evy area drift term affects the GRP rigid body dynamics. We will discuss in particular how the matrix $\mathsf{s}$ is responsible for an additional drift that is analogous to the It\^o-Stratonovich correction, which we will refer to as the L\'evy area correction.

The study of additional drift terms caused by the L\'evy area originated prior to rough paths with the problem of approximation of SDE, and the related Wong-Zakai theorem \cite{Wong1965a, Wong1965b}. One statement of the Wong-Zakai theorem is that smooth approximations of one dimensional SDE converge to the Stratonovich interpretation \cite{Sussmann1978}. The Wong-Zakai theorem often justifies the use of white noise in many stochastic models for unresolved scales, since although correlations may be more accurately modelled by (for example) coloured noise, these correlations are small and constitute a good approximation of the Stratonovich interpretation of a stochastic differential equation. Appealing to the Wong-Zakai theorem is only correct for dimension one,  as counter examples to the Wong-Zakai theorem exist in two dimensions and above \cite{Sussmann1991}, and without additional assumptions a smooth approximation of the SDE may converge to neither the It\^o nor {the} Stratonovich interpretation of the {stochastic integral}. In particular, we shall find here that it may converge to a Stratonovich equation with an additional drift term.
The Wong-Zakai theorem continues to hold in the case where the vector fields defining the multiplicative noise mutually commute, as discussed below.

The modification of the Wong-Zakai theorem that is required by noncommutativity of the vector fields defining the noise in more than one dimension is derived in Ikeda-Watanabe~\cite{Ikeda1981}[Ch7, Thm 7.2]. Namely, let $\{A_n\}_{n=0}^r$ be (smooth) vector fields for a piecewise approximation of Brownian motion, then in the limit that the approximation converges to white noise one obtains an additive drift term modifying the equation which is proportional to the commutator of the smooth vector fields, as
\begin{equation}
\dn X_t = \sum_{n=1}^r A_n(X_t) \circ \dn W_t + A_0(X_t) \dn t + \sum_{1 \leq i \leq j \leq r}s_{ij}[A_i, A_j](X_t) \dn t    \,.
\label{ikedawatanabe}\end{equation}
 
Here, $[\cdot, \cdot]$ denotes the Jacobi-Lie bracket of vector fields, $\circ$ denotes Stratonvich integration and $s_{ij}$ denotes a constant skew-symmetric matrix depending on the nature of the approximation (this is what we shall identify as the L\'evy area perturbation $\mathsf{s}$). Clearly, this additional drift term vanishes when the row vector fields of the diffusion matrix commute. It also vanishes under the assumption of reducibility, for which the Lie brackets vanish, see \cite{Ait-Sahalia-2008}[Prop 1]. Note that these conditions are always satisfied in one-dimensional systems.

In the limit as a smooth approximation of the noise process converges to white noise, multiscale analysis shows that one obtains a stochastic integral that is neither It\^{o}-nor-Stratonovich, at least in the case where the noise is modeled as a stationary Gaussian process with exponential autocorrelation function--see Appendix \ref{sec:colouredappendix} and ~\cite{Pavliotis2008}[Sec. 11.7], \cite{PhysRevE.88.062150}, \cite{Pavliotis2014}{[Ch. 5.1]}. The additional drift term stemming from the resulting neither-It\^{o}-nor-Stratonovich stochastic integral corresponds to the L\`{e}vy area correction. Such additional drift terms can have a dramatic effect on the qualitative properties of the limiting SDE, see~\cite{Pavliotis2008}[Sec. 11.7.7] for a  simple example. One of the questions that we address in this paper is the effect of the L\'{e}vy area correction on the long time behaviour of the limiting dynamics and, in particular, on the invariant measure {for the (dissipative) motion on coadjoint orbits of the Lie-Poisson equations with double bracket selective energy decay as in equation \eqref{Sto-RB-stra-dbrkt}. The answer to this question is given in Theorem \ref{thm: Gibbs} in general terms at the end of section \ref{sec:3}, and it is evaluated computationally in the special case of rigid body dynamics on geometric rough paths in section \ref{sec:4}. Namely, the L\'evy area correction can have a significant effect on the center-of-probability and energy profile of the exponential Gibbs distribution of the invariant measure for geometric mechanics on rough paths.} Thus, the L\'evy area correction can affect the stability properties of the coarse-grained SDE and also its long time behaviour and non-equilibrium steady states.

The introduction of geometric rough paths provides insight into the augmentation of the Wong-Zakai theorem due to adding the L\'{e}vy area correction term -- the last term in Eqn.~\eqref{ikedawatanabe}. We will refer to this phenomenon as the {\it Wong-Zakai anomaly}. Intuitively, one can think of a rough path $(B, \mathbb{B})$ as an ordered pair comprising the path itself and its iterated integrals expressed as a 2-tensor. Even though the path may converge uniformly to Brownian motion, its iterated integrals may fail to converge to the correct object. In many such cases, the symmetric part of the area enhancement converges correctly to the area defined with Stratonovich integration, but the antisymmetric L\'evy area {\color{black}converges with an additional deterministic, antisymmetric correction term}. Thus, the L\'evy stochastic area correction appears as an additional drift term when the corresponding SDEs are interpreted as rough differential equations. By accounting for L\'evy area corrections we can once again invoke the Wong-Zakai theorem to justify modelling unresolved scales as white noise, the difference now being the use of a more general extension of Wong-Zakai as discussed in \cite{Ikeda1981, Sussmann1978, Gyongy2004, Mackevicius1985} and explained via rough paths.

The advantage of rough paths is that they provide pathwise information, rather than the ensemble information obtained from stochastic paths via many individual pathwise realisations. The theory of (geometric) rough paths provides us with a natural framework for studying limit theorems for SDEs, including doing multiscale analysis and understanding the Wong-Zakai theorem and its extensions. In applications, the L\'evy area connection appears naturally when considering the problem of Brownian motion in a magnetic field~\cite{Friz_2015}, as well as in extensions of the Wong-Zakai theorem~\cite{Hairer2020, Hairer2015a, Bass_al_2002}. In this paper, we study  geometric mechanics on rough paths. The theory of rough paths is well suited for studying dynamics with an underlying geometric structure, as already discussed in the two companion papers \cite{Leahy2020,Crisan2022}. We also mention that the L\'evy area correction and its analysis using geometric rough paths plays an important role when studying inference problems for rough SDEs~\cite{Diehl_al_2016, coghi2021robust}. Inference of multiscale stochastic geometric mechanics is an important problem to which we expect to return in future work.



\paragraph{Main objectives of this paper.} This paper investigates the effect of the L\'evy area {correction} both theoretically and numerically for stochastic rigid body dynamics. 

After this introduction, the paper is organised in three more sections and an appendix, as follows. 

\begin{itemize}
  \item In Section \ref{sec:2} we define a rough Hamilton-Pontryagin variational principle in the sense of \cite{Leahy2020}, where the L\'evy area correction is included in the Lagrange multiplier. We show that the resulting equations from the variational principle in Section \ref{sec:2} are equivalent to an Euler-Poincar\'e system driven by a rough path with non-canonical L\'evy area.
  \item In Section \ref{sec:3} we examine the GRP dynamics of Lie-Poisson equations with double-bracket coadjoint dissipation.  We derive the Fokker-Planck generator for these equations and its corresponding stationary distribution for the rough path dynamics. This derivation implies the results summarised in Theorem \ref{thm: Gibbs} which states formally that the invariant measure will be Gibbsian for a wide class of Lie-Poisson dynamics with double-bracket dissipation and L\'evy-area correction arising in the class of rough Hamilton-Pontryagin variational principles posed in Section \ref{sec:2}.
  \item The theoretical results in Sections \ref{sec:2} and \ref{sec:3} are demonstrated in Section \ref{sec:4} with simulations of GRP rigid body dynamics with double-bracket coadjoint energy dissipation. The rigid body is a key example in geometric mechanics. We also show in Section \ref{sec:4} that the theory of non-canonical L\'evy area correction developed in the preceding sections can arise from the well-known approximation problem of "physical Brownian motion" for stochastic equations.
  \item In the appendix \ref{sec:Appendix} we discuss the key results and theorems about GRP that we apply in our primary example of rigid body dynamics driven by coloured Gaussian noise.
\end{itemize}


\section{The L\'evy Area Correction for Variational Principles}\label{sec:2}

In this section we will briefly discuss the motivation, notation and background for our investigation.

\paragraph{Background.} Variational principles on geometric rough paths were introduced in \cite{Leahy2020} and applied to derive rough equations for continuum fluid dynamics. By connecting the theory of rough paths to geometric mechanics the question of how the L\'evy area of the rough path impacts the resulting rough Euler-Poincar\'e equations has become relevant in the mechanics context. As shown in \cite{Friz2009}, we can always choose to work with rough paths enhanced with the canonical choice of L\'evy area, $\mathsf{s} = 0$. Non-canonical L\'evy areas will produce a correction term. Next we will introduce a variational principle on geometric rough paths that involves a non-canonical choice of the L\'evy area. This non-canonical choice will enable us to compare coloured noise models for the SALT approach for rigid body dynamics to the previous Brownian noise approach via SALT in \cite{Arnaudon2018}. 


Throughout this paper, the symbol $\mb{Z}_{s,t} = (Z_t - Z_s, \mathbb{Z}_{s,t}) \in \mc{C}^\alpha_T(\mathbb{R}^d)$ will denote a geometric rough path with H\"{o}lder index $\alpha \in (\frac{1}{3}, \frac{1}{2})$ for all times $t$ in some interval $[0,T]$, $T > 0$. Functions defined on Lie groups, their Lie algebras or other geometric spaces will be assumed to be smooth whilst any time dependence will be stated as being H\"{o}lder continuous for a given exponent unless stated otherwise. 

We will also denote by $$\widetilde{\mb{Z}}_{s,t} = (Z_t - Z_s, \mathbb{Z}_{s,t} + \mathsf{s}(t-s)) \in \mc{C}^\alpha_T(\mathbb{R}^d)$$ a geometric rough path obtained from $\mb{Z}$ by perturbing its L\'evy area, that is, $\mb{Z}$ and $\widetilde{\mb{Z}}$ have the same trace/path component $Z$ yet differ by an antisymmetric matrix $\mathsf{s}$ in their signature tensor. Note that $\mathsf{s}$ must be antisymmetric to maintain the identity $\operatorname{Sym}(\mathbb{Z} + \mathsf{s}(t-s)) = \frac{1}{2}Z \otimes Z$, which ensures that $\widetilde{\mb{Z}}$ is geometric. 
We shall be referring to $\mathsf{s}$ as a perturbation to the L\'evy area, named as such since the case of Stratonovich enhanced Brownian motion the additional term is a determinstic addition to L\'evy's stochastic area.

Let $G$ be an $n$-dimensional matrix Lie group with Lie algebra $\mathfrak{g}$. Suppose $G$ acts (on the left) of a tangent bundle $TQ$ where a physical system is being described by a Lagrangian $L: TQ \rightarrow \mathbb{R}$. When $L$ is (left) invariant under $G$ the dynamics on $TQ$ can be lifted to motion on $TG$ via identification of curves $x_t = g_t x_0$ when the $G$-action is transitive. The invariance of $L$ allows us to study the motion taking place on a reduced space where the coordinates are invariant under symmetry by defining a reduced Lagrangian $l(\xi) := L(x_0, g^{-1}\dot{g_t}x_0)$. The action of $g^{-1}$ on the velocity of the curve pulls back $\dot{g}$ from $T_g G$ to $T_e G = \mathfrak{g}$ and this continues to hold true for stochastic/rough differential equations with differential $\dn g$. It is dynamics on the Lie algebra $\mathfrak{g}$ that we shall be interested in and henceforth we assume this type of reduction procedure has been performed.  

Given a symmetry-reduced Lagrangian $l : \mathfrak{g} \rightarrow \mathbb{R}$, we are interested in deriving equations from Hamilton's principle $\delta S = 0$ with  the following equivalent rough Hamilton-Pontryagin action integrals \cite{Leahy2020,holm2009geometric,holm2011geometric}, 

\begin{equation}S^{\text{HP}}_{\widetilde{\mb{Z}}}(u, g, \mu)=\int l(u) \dn t+\int\left\langle\mu, g^{-1} \dn  g-u \dn t +\xi \dn \widetilde{\mathbf{Z}}\right\rangle\,,
\label{varprinciple}
\end{equation}

\begin{equation}S^{\text{HP}}_{\mathsf{s}, \mb{Z}}(u, g, \mu)=\int l(u) \dn t+\int\left\langle\mu, g^{-1} \dn  g-u \dn t + \frac{1}{2}\mathsf{s}^{ij}[\xi_i, \xi_j] \dn t+\xi \dn \mathbf{Z}\right\rangle\,.
\label{othervarprinciple}
\end{equation}

The quantity $\mu \in \mathfrak{g}^*$ is the Lagrange multiplier enforcing the equivalent constraints
\begin{equation}
g^{-1}\dn g 
=  u\dn t  {\color{black}-} \xi \dn \widetilde{\mb{Z}}     
\,,
\label{velocity}\end{equation}
\begin{equation}
g^{-1}\dn g 
=  u\dn t - \frac{1}{2}\mathsf{s}^{ij}[\xi_i, \xi_j] \dn t {\color{black}-} \xi \dn \mb{Z}     
\, ,
\label{othervelocity}
\end{equation}
respectively. For $g \in G$ and $u, \xi_i \in \mathfrak{g}, 1 \leq i \leq d$. The notation $\xi \dn {\mb{Z}}$ denotes the sum $\sum_i \xi_i \dn {\mb{Z}}^i$, when writing the L\'evy area correction repeated indices will imply summation. 

We will establish in the next section that the constraints \eqref{velocity} and \eqref{othervelocity} both lead to the same Euler Poincaré equations which would have been clearly distinct had we considered no change in the L\'evy area of the driving noise. For the case the rough path is Brownian motion these correspond to stochastic Euler-Poincar\'e equations where the integral is neither It\^o nor Stratonovich. We anticipate that \eqref{varprinciple} is a more natural choice for applications\footnote{For example, high wavenumbers caused by thermal effects in the ocean may warrant use of coloured noise. The white noise limit of the resulting SPDE may lead to a limiting SPDE with a nontrivial L\'{e}vy area correction in the drift.}, since one may observe a rough velocity with $\dn \widetilde{\mb{Z}}$ without determining the tensor $\mathsf{s}$. For the numerical examples in \ref{sec:4} where we investigate the effects of known values of $\mathsf{s}$ using standard (Stratonovich) Brownian motion as our rough path, we will make use of the form in \eqref{othervarprinciple} with the explicit drift term involving $\mathsf{s}$. 

In finite dimensions, the L\'evy area correction may be defined in terms of the commutator of elements $\xi_i=\Xi^{ib}\hat{e}_b$ of a matrix Lie algebra $\mathfrak{g}$ with structure constants $c_{ab}^c$ given in terms of the basis $\hat{e}_a$ with $a=1,2,\dots,{\rm dim}(\mathfrak{g})$ by $[\hat{e}_a,\hat{e}_b]=c_{ab}^c\hat{e}_c$. The quantity $[\xi_i, \xi_j]=\Xi^{a}_i\Xi^{b}_jc_{ab}^c\hat{e}_c$ is always a left-invariant matrix Lie algebra element. Thus the L\'evy area correction can be identified as a Lie algebra element formed from the bracket in the matrix Lie algebra $\mathfrak{g}$.
Hence, in this paper we will generally take the $\xi_i$ as fixed matrix Lie algebra elements.

Noise where $\xi_i$ possess functional dependence on certain dynamical quantities such as $\mu$ is possible when formulated used Clebsch constraints \cite{Arnaudon2018}, but will not be considered in this paper. In the SALT approach the resulting rough differential equation (RDE) obtained from stationarity of the action integral in \eqref{varprinciple} will have a multiplicative noise due to the coadjoint representation and momentum map terms appearing in the resulting Euler-Poincar\'e equation. The fact that constraint \eqref{velocity} has additive noise in the case of the SALT approach can be useful for designing numerical methods for the Lie-Poisson equations, as discussed in \cite{Luesink2021}.  
\paragraph{Intepretation of the variational principle.} Constraints \eqref{velocity} and \eqref{othervelocity} imposed by the Lagrange multiplier $\mu$ in the variational principles \eqref{varprinciple} and \eqref{othervarprinciple}, respectively, define two representations of the same transport velocity. The next section will discuss these two representations of the transport velocity in detail.

\paragraph{Links to Wong-Zakai anomalies and approximation of SDE.} From the rough paths perspective of \cite{Friz2009}, the additional drift term $\frac{1}{2}\mathsf{s}^{ij}[\xi_i, \xi_j]$ in \eqref{othervelocity} arises when non-canonical choices of iterated integrals are made for the driving noise.  Representing a change of Galilean frame, this drift term is analogous to the quadratic covariation correction term that arises when converting between It\^o and Stratonovich interpretations of white noise \cite{Holm2020}. Correction terms of the type $\frac{1}{2}\mathsf{s}^{ij}[\xi_i, \xi_j]$ arise in both SDE analysis, see e.g., \cite{Sussmann1991,Ikeda1981,Gyongy2004}, as well as in rough paths theory \cite{Friz2009}. 

\subsection{Rough Euler-Poincar\'e equations}

The rough variational principle in \eqref{varprinciple} can be interpreted as a rough integral, \cite{Leahy2020}. Critical points of the action principle \eqref{varprinciple} produces a rough Euler-Poincar\'e equation in which the rough driver $\dn \widetilde{\mb{Z}}$ contains a modification of its L\'evy area. Here, we will show this by rough variational principle yields the following distinct Euler-Poincar\'e equation with an additional `drift term', cf. \eqref{othervelocity},

\begin{equation}
    \dd \mu - \ad^*_{(u\dn t - \frac{1}{2}\mathsf{s}^{ij}[\xi_i, \xi_j] \dn t - \xi \dn \mb{Z}) }\mu = 0,
\label{eplevy}
\end{equation}
where the coadjoint operator $\ad^*$ is defined as the dual of the map $\ad_\xi \eta :=[\xi,\eta]$, satisfying  \cite{holm2009geometric, holm2011geometric}
\begin{align}
\begin{split}
\mathrm{ad}^{*}: \mathfrak{g} \times \mathfrak{g}^{*} & \longrightarrow \mathfrak{g}^{*}, \\
(\xi, \mu) & \mapsto \operatorname{ad}_{\xi}^{*}(\mu),\\
 \left\langle\operatorname{ad}_{\xi}^{*} \mu, \eta\right\rangle & =\left\langle\mu, \operatorname{ad}_{\xi} \eta\right\rangle, \quad \text { for all } \eta \in \mathfrak{g} .
\end{split}
\label{def: ad-star}
\end{align}
Linearity of the adjoint action allows us to include the L\'evy area correction in \eqref{eplevy} as part of the mean field drift velocity contribution $\ad^*_u \mu \dd t$, providing an interpretation as a change in Galilean frame from velocity $u$ to relative velocity $u - \frac{1}{2}\mathsf{s}^{ij}[\xi_i,\xi_j]$.
We note that the sign of the $\ad^*$ operator in equation \eqref{eplevy} depends on whether the Lagrangian in variational principles \eqref{varprinciple}, \eqref{othervarprinciple} is left or right invariant, see equations \eqref{eqn: SALTEuler} and \eqref{Sto-RB-SALT} for example. Throughout this paper the abstract Euler-Poincare and Lie-Poisson equations are written for left invariance. For right invariance as in the case of fluid dynamics, one has $\ad_\xi \eta := -[\xi,\eta]$ and the sign in front of the $\ad^*$ operator in \eqref{eplevy} is positive. 

The main result of this section is the following proposition.

\begin{proposition}{For the same initial data, solutions of the following two Euler-Poincar\'e equations coincide }
\begin{equation}
    \dd \mu + \ad^*_u \mu \dd t {\color{black}-} \frac{1}{2}\mathsf{s}^{ij}\ad^*_{[\xi_i, \xi_j]} \mu\dd t + \ad^*_{\xi}\mu \dd \mb{Z} = 0
    \,,\label{EPa}
\end{equation}
\begin{equation}
    \dd \mu + \ad^*_{u}\mu \dd t + \ad^*_\xi \mu \dd \widetilde{\mb{Z}} = 0\,.\label{EPb}
\end{equation}

\label{rdeproposition}\end{proposition}

\textbf{Proof:} Let $\mu : [0,T] \rightarrow \mathfrak{g}^*$ be a solution to the rough Euler--Poincar\'e equations (with no correction term) driven by the rough path $\widetilde{\mb{Z}}$. That is, $\mu$ solves the RDE

\begin{equation}
    \dd \mu + \ad^*_{u}\mu \dd t + \ad^*_\xi \mu \dd \widetilde{\mb{Z}} = 0.
\label{eptilde}\end{equation}

Solving this RDE is equivalent to the following asymptotic estimate:

\begin{equation}
    \delta\mu_{st} + \int_{s}^{t}\ad^*_{u_r}\mu_r \dd r + \ad^*_{\xi_k} \mu_s \delta Z^k_{st} + \left(\ad^*_\xi \mu_s\right)^\prime\widetilde{\mathbb{Z}} = o\left(|t-s|\right).
\label{integralform}\end{equation}

Here the $\delta$ operator (not to be confused with variations elsewhere in this paper) is shorthand for $\delta f_{st} := f_{t}-f_{s}$ with $f \in C_{T}^{\alpha}(\mathbb{R}^d)$ and prime $(\prime)$ denotes taking the Gubinelli derivative \cite{Gubinelli2004}. A standard property of the Gubinelli derivative is that $F(\mu)^\prime = DF(\mu)\mu^\prime$ for any (Frechet) differential function, $F$. Furthermore, since $\mu$ solves \eqref{eptilde} we may conclude that $\mu^\prime = \ad^*_\xi \mu$. Since $\widetilde{\mathbb{Z}} = \mathbb{Z} + \mathsf{s}(t-s)$ we may then rewrite \eqref{integralform} as follows

\begin{equation}
    \delta\mu_{st} + \int_{s}^{t}\ad^*_{u_r}\mu_r \dd r + \ad^*_{\xi_k} \mu_s \delta Z^k_{st} + \left(\ad^*_\xi \mu_s\right)^\prime\mathbb{Z} + \ad^*_{\xi_i} \ad^*_{\xi_j} \mu_s\mathsf{s}^{ij}(t-s) = o\left(|t-s|\right)
\,.\end{equation}

Antisymmetry of $\mathsf{s}$ allows us to write $\mathsf{s}^{ij} = \frac{1}{2}\left(\mathsf{s}^{ij} - \mathsf{s}^{ji}\right)$. Relabelling indices then allows us to write the final term as \[\frac{1}{2}\mathsf{s}^{ij}\left(\ad^*_{\xi_i} \ad^*_{\xi_j}\mu_s - \ad^*_{\xi_j} \ad^*_{\xi_i} \mu_s \right)(t-s) = {\color{black}-}\frac{1}{2}\mathsf{s}^{ij}\ad^*_{[\xi_i, \xi_j]}\mu_s(t-s)\,.\] Use of the standard estimate $\int_s^tY_{r}\dn X_r = Y_s\delta X_{st} + o(|t-s|)$ finally allows us to conclude

\begin{equation}
    \delta\mu_{st} + \int_{s}^{t}\ad^*_{u_r}\mu_r \dd r {\color{black}-} \int_{s}^{t}\frac{1}{2}\mathsf{s}^{ij}\ad^*_{[\xi_i,\xi_j]}\mu_r \dd r+ \ad^*_{\xi_k} \mu_s \delta Z^k_{st} + \left(\ad^*_\xi \mu_s\right)^\prime{\mathbb{Z}} = o\left(|t-s|\right)
\,.\end{equation}

This is the Davie formulation for $\mu$ to be a solution of the RDE \eqref{eplevy}. Consequently, solutions of the two rough differential equations \eqref{EPa} and \eqref{EPb} coincide, upon invoking the uniqueness property of RDE \cite{Hairer2020}.

\subsection{Rough Lagrange-d'Alembert Variational Principle}\label{2.2}

{\color{black}In the previous section we discussed the variational principle \eqref{varprinciple} introduced in \cite{Leahy2020} and defined it for rough paths with modified L\'evy area. The variational principle \eqref{varprinciple} results in Lie-Poisson equations for coadjoint motion with {modified drift}. The effects of the L\'evy area correction term on the solutions of the double-bracket dissipative Lie-Poisson equations are of interest to us, because in the stochastic Lie-Poisson case the { invariant measure is Gibbsian}. The aim of this paper is to find out whether the Gibbs property persists in the presence of L\'evy area correction terms. In order to derive {the} rough dissipative Lie-Poisson equation we will formulate a rough paths generalisation of the well known Lagrange d'Alembert variational principle.} 

Stochastic generalisations of the Lagrange-d'Alembert approach for introducing constraints into variational principles were introduced in \cite{Hu2021, Kraus2020}. It will be useful to formulate the Lagrange-d'Alembert approach {using rough integrals as in}  \eqref{varprinciple} in order to {make possible} the addition of double-bracket dissipation to the rigid body equations recast as a RDE in Section \ref{sec:4}. Let $S^{\text{HP}}_{\mathsf{s}, \mb{Z}}$ denote the action as defined in the previous section. Let $F_0, F^i : T^*M \rightarrow \mathfrak{g}^*$ with $F_0 \in C^\alpha_T$ for $\alpha \in (\frac{1}{2}, \infty)$ and $F^i \in \mathcal{D}_{\mb{Z}, T}$. However, for our purposes in Sections \ref{sec:3} and \ref{sec:4} we shall make the simplifying assumption that $F^i = F^i(\mu)$ and $F_0 = F_0(\mu)$ are smooth functions of the momentum map. We consider the following variational principle

\begin{equation}
\delta S^{\text{HP}}_{\mathsf{s}, \mb{Z}} - \int \langle F^0 , \delta g g^{-1} \rangle \dn t 
-  \int \langle F^i , \delta g g^{-1} \rangle \dn \mb{Z}^i = 0\,.
\label{lagrangedalembert}\end{equation}

{This leads to} the following forced Euler-Poincar\'e equations

\begin{equation}
    \dd \mu + \left(\ad^*_u \mu + F^0\right)\dn t {\color{black}-} \frac{1}{2}\mathsf{s}^{ij}\left( \ad^*_{[\xi_i, \xi_j]} \mu \right)\dn t + \left(\ad^*_{\xi_i}\mu + F^i\right) \dn \mb{Z}^i = 0\,.
\label{forcedeulerpoincare}\end{equation}

The proof is standard, except that one takes variations $\delta g \in C^\infty_T(G)$ with the first integral in \eqref{lagrangedalembert} interpreted in the Young \cite{Young1936} sense, and the second integral interpreted as a rough integral of a continuous bilinear form $\langle F^i , \delta g g^{-1}\rangle$ evaluated at two controlled rough paths (see Remark A.9 of \cite{Leahy2020}). { All examples of the stochastic rigid body dynamics appearing in the remainder of this paper can be obtained as special cases of \eqref{forcedeulerpoincare}.}

\subsection{A note on Hamiltonian reduction and modified drift terms}\label{2.3}


{We remark that the L\'evy area correction is compatible with the process of Hamiltonian reduction, in the sense that applying the same procedure as Proposition \ref{maintheorem} for a rough symplectic vector field in phase space and reducing via symmetry produces the associated Lie-Poisson equations corresponding to equation \eqref{eplevy}.} 

{\color{black} Equation \eqref{eplevy} implies a Lie-Poisson bracket formulation for { the} evolution of observables $f(\mu)$ through the identity $\dn f(\mu) = \langle \frac{\delta f}{\delta \mu}, \dn \mu \rangle$. Conversely, for a  given symplectic form $\omega$, it is possible to recover Lie-Poisson equations provided the Hamiltonian collectivises. Let us explain this comment in what follows.

Abstractly, one can start with a symplectic  manifold $(P, \omega)$ and reduce its symplectic Poisson bracket structure $\{\cdot, \cdot\}_\omega$ induced by $\omega$. A symplectic form $\omega$ allows the definition of Hamiltonian vector fields $\mathbf{d} f = \iota_{X_{f}} \omega$, where $\mathbf{d}$ denotes the exterior derivative.

}  
Given a collection of left-invariant functions $\mathcal{H}_i : T^*G \rightarrow \mathbb{R}$, with $\mathcal{H}_0$ the Hamiltonian and potentials $\mathcal{H}_i, 1 \leq i \leq d$, consider the rough vector field defined by the following RDE

\begin{equation}
\mathrm{d} Y=X_{\mathcal{H}_{0}}\left(Y\right) \mathrm{d} t+\sum_{k=1}^{N} X_{\mathcal{H}_{k}}\left(Y\right) \dn \widetilde{\mb{Z}}^{k}, \text { where } \mathbf{d} \mathcal{H}_{k}= \iota_{X_{\mathcal{H}_{k}}} \omega
\,,\label{roughsymplectic}
\end{equation}
in which $\iota_{X} \omega$ denotes insertion of the vector field $X$ into the symplectic 2-form $\omega$.

The corresponding Poisson bracket formulation is given by

\begin{equation}\mathrm{d} f\left(Y\right)=\left\{f, \mathcal{H}_{0}\right\}_\omega\left(Y\right) \mathrm{d} t+\sum_{k=1}^{N}\left\{f, \mathcal{H}_{k}\right\}_\omega \left(Y\right) \mathrm{d} \widetilde{\mb{Z}}^{k}
\,.\label{roughpoissonbracket}
\end{equation}

In the case that $\tilde{\mb{Z}}$ is a rough path with non-canonical L\'evy area such that $\widetilde{\mathbb{Z}}_{s,t} = \mathbb{Z}_{s,t} + \mathsf{s}(t-s)$, then we have a L\'evy area correction involving Lie bracket of the vector fields $X_{\mathcal{H}_{i}}$ (see appendix Theorem \ref{friztheorem}). Use of the identity 
\[
[X_{\mathcal{H}_{i}}, X_{\mathcal{H}_{j}}] = {\color{black}-} X_{\{\mathcal{H}_{i} , \mathcal{H}_{j} \}_\omega}
\] 
results in the following rough symplectic vector field:
\begin{equation}
\mathrm{d} Y=X_{\mathcal{H}_{0}}\left(Y\right) \mathrm{d} t \,{\color{black}-} \sum_{ij}X_{\frac{1}{2}\mathsf{s}^{ij}\{\mathcal{H}_i, \mathcal{H}_j\}_\omega}(Y) \dn t  +\sum_{k=1}^{N} X_{\mathcal{H}_{k}}\left(Y\right) \dn \mb{Z}^{k}
\,.
\label{hamiltonlevy}
\end{equation}

Note that \eqref{hamiltonlevy} continues to preserve the symplectic form as a consequence of the Lie derivative identity 
\[
\pounds_{X_{\{f, g\}_\omega}}= {\color{black}-}\pounds_{[X_f, X_g]} 
= {\color{black}-}\Big(\pounds_{X_f}\pounds_{X_g} - \pounds_{X_g}\pounds_{X_f}\Big)
\,.\]
The existence of a cotangent lift right momentum map $J_R : \mathfrak{g}^* \rightarrow T^*G$ for the Poisson bracket defined in \eqref{roughpoissonbracket} allows us to define a reduced Hamiltonian $\mathcal{H}_0 = h \circ J_R $ and potentials on the Lie algebra $\mathcal{H}_i = \xi_i \circ J_R$. The newly defined reduced Hamiltonians and potentials allow writing the Lie-Poisson equation, whose Poisson bracket formulation is written using the Lie-Poisson bracket $\{\cdot, \cdot \}$. The L\'evy area correction term is compatible with this procedure because the cotangent lift momentum map is equivariant, 
\[
\{ \mathcal{H}_i, \mathcal{H}_j \}_\omega = \{\xi_i \circ J_R, \xi_j \circ J_R\}_\omega = \{\xi_i, \xi_j\} \circ J_R
\,.\]
Consequently, equation \eqref{hamiltonlevy} implies a set of Lie-Poisson equations 

\begin{equation}
    \dn f = \left\{f, h {\color{black}-} \frac{1}{2}\mathsf{s}^{ij}\{\xi_i, \xi_j\}\right\}\dn t + \sum_{k=1}^d \{f, \xi_k\} \dn \mb{Z}^k\,.
\end{equation}

These are the Lie-Poisson equations obtained when a L\'evy area correction is included.

\section{Application to Stochastic Geometric Mechanics}\label{sec:3}

This section studies the stochastic Lie-Poisson equation with an additional drift caused by the L\'evy area. This example can be treated as the special case in which our rough path $\mb{X}$ is Stratonovich Brownian motion, with signature tensor defined through Stratonovich integrals $\mathbb{B}^{\text{Strat}}_{s,t} := \int_s^t B^i_r \circ \dn B^j_r$ for some Brownian motion $B_t$. The geometric structure of the stochastic Lie-Poisson equations and their infinitesimal generator has already been studied in \cite{Arnaudon2018}.  We demonstrate that the L\'evy area correction term for SALT \cite{Holm2015, Luesink2021} (in which $\xi_i$ are specified Lie algebra elements) also maintains this geometric structure. The double $\operatorname{ad}^*$ dissipation we consider in this section will modify the minimum energy state appearing in the resulting Gibbs measure.

Throughout this section we assume that the Lie algebra $\mathfrak{g}$ is finite dimensional in order to ensure we consider invariant measures of SDE and avoid discussions of invariant measures of stochastic partial differential equations. The finite dimensional assumption also allows use of the structure constants of the Lie algebra, and we will further assume that $\mathfrak{g}$ is semisimple in order to guarantee the existence of a bi-invariant Killing form, as well as the divergence free property of $\ad^*_{\mu^\sharp}(\cdot)$ and $\ad^*_{(\cdot)}\mu$. \footnote{This fact is demonstrated by {$\operatorname{div}_\mu \ad^*_{\mu^\sharp}  = -\frac{\partial}{\partial \mu^k}c^k_{ij} B^{lj}{\mu}_l = -B^{kj}c^k_{ij}  = 0$ where $(B^{ij})^{-1} = c_{im}^n c_{jn}^m $ denotes the components of the inverse matrix of the Killing form $B = B^{ij} \mathbf{\hat{e}}_i \otimes \mathbf{\hat{e}}_j$ and the last equality follows from vanishing of the contraction of a symmetric and antisymmetric tensor. Similar arguments hold for $\ad^*_{\mu^\sharp}F(\mu)$ if $\frac{\partial F^i}{\partial \mu^k}$ is symmetric in its indices.}}  

\subsection{Invariant Measures}

In order to calculate the infinitesimal generator of the stochastic Lie-Poisson equations we must first transform to It\^o noise. We shall reproduce the proof in \cite{Arnaudon2018} and \cite{Holm2015} for convenience, and see that the It\^o-Stratonovich correction manifests as a double $\ad^*$ term. Consider the SDE $\dn q = \ad^*_{\xi_i} q \circ \dn W^i$, we can pair this with a time independent element of the Lie algebra $\phi$ allowing us to say $\dn  \langle \phi, q \rangle = \langle \phi, \dn q \rangle $, it follows that:

$$\dn  \langle \phi, q \rangle = \langle \phi, \ad^*_{\xi_i} q \circ \dn W^i \rangle = \langle \ad_{\xi_i} \phi, q \rangle \circ \dn W^i = \langle \ad_{\xi_i} \phi, q \rangle\dn W^i + \frac{1}{2}[\dn \langle \ad_{\xi_i} \phi, q \rangle, \dn W^i] $$

$$ = \langle \ad_{\xi_i} \phi, q \rangle\dn W^i + \frac{1}{2}[\langle \ad_{\xi_j}\ad_{\xi_i} \phi, q \rangle \dn W^j, \dn W^i] = \langle \phi, \ad^*_{\xi_i} q \rangle\dn W^i + \frac{1}{2}\langle \phi, \ad^*_{\xi_i}\ad^*_{\xi_i} q \rangle$$

We obtain the It\^o form of the Lie-Poisson equations, which will allow us to read off the form of the generator from the drift terms, as:

\begin{equation}
    \dn \mu = \ad^*_{\frac{\partial h}{\partial \mu}}\mu \dn t + \ad^*_{\xi}\mu \dd B - \sum_{1 \leq i \leq d}\frac{1}{2}\ad^*_{\xi_i} \ad^*_{\xi_i}\mu \dn t. 
\label{itoform}\end{equation}

Where the (reduced) Hamiltonian $h: \mathfrak{g}^* \rightarrow \mathbb{R}$ is such that $\frac{\partial h}{\partial \mu} = u \in \mathfrak{g}$. In order to write the generator we make use of the Lie-Poisson bracket. We define the Lie-Poisson bracket by \[\{f,g\}(\mu) := \left\langle \left[\frac{\partial f}{\partial \mu}, \frac{\partial g}{\partial \mu}\right], \mu \right\rangle, \] on functions $f,g: \mathfrak{g}^* \rightarrow \mathbb{R}$ with a derivative $\frac{\partial f}{\partial \mu}, \frac{\partial g}{\partial \mu}: \mathfrak{g}^* \rightarrow \mathfrak{g}$. The expressions $[\,\cdot\,,\, \cdot\,]$ and $\langle \,\cdot\,,\, \cdot\, \rangle$ are the Lie algebra bracket and dual pairing for $\mathfrak{g}$, respectively. In the SALT approach used here, we can consider the noise amplitude $\xi_i$ as arising from a stochastic potential $\Phi_i(\mu) = \langle \xi_i, \mu \rangle$ which has derivative $\frac{\partial \Phi_i}{\partial \mu}$ equal to $\xi_i$, so we can write the generator of \eqref{itoform} using Poisson brackets of $\Phi$. The Poisson bracket formulation of the Lie-Poisson equations readily allows calculation of the infinitesimal generator by use of the identity $\dn f(\mu) = \left\langle \frac{\partial f}{\partial \mu}, \dn \mu\right\rangle$. Upon substituting in \eqref{itoform} and using $\ad_x := [x, \cdot]$, we immediately conclude the form of the generator:

\begin{equation}
    \mathcal{L}f = \{h, f \}  + \sum_{1 \leq i \leq d}\frac{1}{2}\{\Phi_i, \{\Phi_i, f \}\}.
\label{itogen}\end{equation}

Considering an additional drift caused by the L\'evy area means we must include an additional term in $\mathcal{L}$, determined in the same way as for obtaining \eqref{itogen}. The evolution of a function of the momentum $f(\mu)$ is calculated as 

$$\dn f(\mu) = \left\langle \frac{\partial f}{\partial \mu}, \dn \mu\right\rangle = \left\langle \frac{\partial f}{\partial \mu}, \ad^*_{\frac{\partial h}{\partial \mu}} \mu \dd t {\color{black}-} \frac{1}{2}\mathsf{s}^{ij}\ad^*_{[\xi_i, \xi_j]} \mu\dd t + \ad^*_{\xi}\mu \circ \dn B\right\rangle.$$

The L\'evy area induces a contribution of ${\color{black}-}\left\langle \frac{\partial f}{\partial \mu}, \frac{1}{2}\mathsf{s}^{ij}\ad^*_{[\xi_i, \xi_j]} \mu \right\rangle$ to the Lie-Poisson equations. We can write a constant Lie algebra element as a variational derivative of a pairing with $\mu$ to write the Lie bracket as a Poisson bracket, along with properties of the coadjoint action this allows us to deduce 

\[
\left\langle \left[ \frac{\partial f}{\partial \mu}, \frac{1}{2}\mathsf{s}^{ij}[\xi_i, \xi_j]\right],  \mu \right\rangle = \left\{f, \frac{1}{2}\mathsf{s}^{ij}\{\Phi_i, \Phi_j\}\right\}.
\]

The generator can thus be written as 

\begin{equation}
    \mathcal{L}_{\mathsf{s}}f = \{h {\color{black}-} \frac{1}{2}\mathsf{s}^{ij}\{\Phi_i, \Phi_j\}, f \}  + \sum_{1 \leq i \leq d}\frac{1}{2}\{\Phi_i, \{\Phi_i, f \}\}.
\label{3.3}\end{equation}

The operator in \eqref{3.3} $\mathcal{L}_\mathsf{s}$ can be written in Hormander's sum of squares form $\mathcal{A} + \mathcal{B}_i^2$, where $\mathcal{A}$ and $\mathcal{B}_i$ are Poisson brackets evaluated in one argument. In addition, the generator is hypocoercive in the sense of Villani. In particular, the operator $\mathcal{A}$ is antisymmetric in (the flat space) $L^2$, provided we have vanishing boundary conditions on the coadjoint orbits (such as on the sphere), and their adjoints are computed using the derivation property of Poisson brackets. The fact that the Kirillov-Kostant-Souriau symplectic form is an exact differential on coadjoint orbits ensures that left over terms vanish by the Stokes theorem. 
This reasoning implies 
\begin{equation}
\mathcal{L}^*_{\mathsf{s}}f = -\{h {\color{black}-} \frac{1}{2}\mathsf{s}^{ij}\{\Phi_i, \Phi_j\}, f \}  + \sum_{1 \leq i \leq d}\frac{1}{2}\{\Phi_i, \{\Phi_i, f \}\}.
\label{3.4}\end{equation}
We remind the reader that a measure $\mu$ is invariant under the dynamics generated by~\eqref{itogen} provided that it is invariant under the action of the adjoint semigroup, $P_t^* \mu = \mu$, see~\cite{Pavliotis2014}[Ch. 2]. Assuming that the invariant measure has a smooth density with respect to the Lebesgue measure,  the invariant density is in the null space of the Fokker-Planck operator $\mathcal{L}^*_{\mathsf{s}}$.

\begin{proposition} For a fixed, spanning $\xi_i \in \mathfrak{g}$ the uniform invariant measure of the Lie-Poisson equations is unchanged when adding the L\'evy area correction.
\label{maintheorem}\end{proposition}


We can now apply a lemma used in \cite{Arnaudon2018}, that the kernel of operator of the form  $\mathcal{L}^*_{\mathsf{s}} = \mathcal{A} + \mathcal{B}_i^2$ is the intersection of their kernels of the individual operators $\mathcal{A}, \mathcal{B}_i$ \cite{Villani2009}.

For our case, $\mathcal{A}$ and $\mathcal{B}_i$ are Poisson operators. By definition, the only element in both kernels are elements which Poisson commute with all functions, these are precisely the Casimirs, which have the property of being constant on the coadjoint orbits, which solutions are necessarily confined to. As a result the kernel of $\mathcal{L}^*_{\mathsf{s}}$ comprises of constant functions, which is in agreement with the result in \cite{Arnaudon2018}. The constant is determined by the volume of the coadjoint orbit the solution lives on. Because we know the L\'evy area correction term can be grouped with the velocity as in \eqref{othervelocity}, the coadjoint motion property is left preserved, this means the solution lives on the same coadjoint orbit with or without the L\'evy area correction, and the value of the constant in the invariant measure is the same.

Thus, the invariant measure for SALT is unchanged by the L\'evy area correction.

We assume spanning $\xi_i$ to ensure geometric ergodicity of the SDE \cite{Mattingly2001} and avoid technical difficulties; in practice we can relax this assumption in computations and observe the same conclusions.

\paragraph{Remark.} In Section \ref{2.3}, we explained heuristically why the L\'evy area correction is compatible with much of the structure of SALT. We note that the L\'evy area correction is similarly well behaved from a stochastic analysis perspective. We now show that the additional drift arising form the L\'evy area can be removed through a change of measure. Girsanov's theorem provides an intuitive argument that the invariant measure must be unaffected. To apply Girsanov's theorem it is sufficient to find a process $u(t, \omega)$ that satisfies the Novikov condition \cite{Oksendal1998}:

$$\mathbb{E} \left[\exp \left(\frac{1}{2} \int_{0}^{T} u^{2}(s, \omega) d s\right)\right]<\infty.$$

The multiplicative noise matrix multiplying $u$ should also equal the L\'evy area drift we wish to try and remove, that is, $u$ is required to solve the linear equation:
\begin{equation}
\ad^*_{\xi_k}\mu u^k = \mathsf{s}^{ij}\ad^*_{[\xi_i, \xi_j]}\mu\,.
\label{Novikov}
\end{equation}

Equation \eqref{Novikov} is an underdetermined linear system. Even if the Lie algebra elements $\xi_i$ comprise a basis of the Lie algebra, the matrix whose $k$-th column is given by $\ad^*_{\xi_k}\mu$ will always have a nontrivial kernel. Nevertheless, there are situations when equation \eqref{Novikov} can be solved, as it is equivalent to finding $u^k$ that satisfy
\begin{equation} 
u^k \xi_k = \mathsf{s}^{ij}[\xi_i, \xi_j] = \mathsf{s}^{ij}c^k_{ij}\xi_k
\quad\Longrightarrow\quad u^k=\mathsf{s}^{ij}c^k_{ij}\,.
\label{LieDepXis}
\end{equation}
In the finite-dimensional case, it is clear that constant coefficients depending on the L\'evy area and structure constants of the matrix Lie algebra can be chosen, since the Lie brackets $[\xi_i, \xi_j]$ depend linearly on the $\xi_i$. 
Since $u^k$ are constants and have no dependence on the stochastic process $\mu$, the Novikov integrability condition holds in this case, and the law of \eqref{eplevy} is equivalent to that of the standard stochastic Lie-Poisson equations.

\subsection{Double-Bracket Dissipation (DBD)}

Double-bracket dissipation (DBD) reviewed in \cite{GayBalmaz2014} introduces energy dissipation that preserves coadjoint motion. DBD is applied to coadjoint stochastic rigid body dynamics in \cite{Arnaudon2018}, as the following modification of the Lie-Poisson equation \eqref{EPb},
\begin{equation}\dn \mu+\operatorname{ad}_{\frac{\partial h}{\partial \mu}}^{*} \mu \mathrm{d} t+\vartheta \operatorname{ad}_{\frac{\partial C}{\partial \mu}}^{*}\left[\frac{\partial C}{\partial \mu}, \frac{\partial h}{\partial \mu}\right]^{\flat} \mathrm{~d} t
+ \sum_{i} \mathrm{ad}_{\xi_i}^{*} \mu \circ \dn W^i_{t}=0.
\label{dissipativeLP}\end{equation}

The solutions of the DBD modified equation \eqref{dissipativeLP} preserve level sets of the Casimir function $C(\mu) = \frac{1}{2}\|\mu\|^2$ and dissipate energy $h(\mu)$ at a rate controlled by the parameter $\vartheta$. Upon modifying the Lie-Poisson equation \eqref{EPb} by including DBD, the invariant measure no longer tends to a constant. Instead when the noise is isotropic $\xi_i = \sigma \mathbf{\hat{e}}_i$, it tends to a Gibbs measure of the form $\mathbb{P} = Z^{-1}\exp(-\frac{2\vartheta }{\sigma^2 }h(\mu))$, where $Z$ denotes the normalisation constant $Z$. The Gibbs measure is derived in \cite{Arnaudon2018} by taking $\mathbb{P} = f(h)$ as an ansatz to cancel out the drift term contribution, $-\{h,\mathbb{P}\}$, in the Fokker-Planck operator and balancing the contribution of the It\^o-Stratonovich correction against the new double-bracket dissipation term. 


The approach in \cite{Arnaudon2018} must be adjusted here to ensure cancellation of both the Hamiltonian and L\'evy area contributions for \eqref{3.3}. We consider an ansatz of the form $\mathbb{P} = f(\widetilde{h})$, {\color{black}in which the modified Hamiltonian \begin{equation}
\widetilde{h} := h -\frac{1}{2}\mathsf{s}^{ij}\{\Phi_i, \Phi_j\} \equiv h -\big\langle  \mu\,,\, \frac12\mathsf{s}^{ij}{[\xi_i, \xi_j]} \big\rangle 
\label{modifiedhamiltonian}\end{equation}
includes the contribution from the L\'evy area correction.} To ensure the DBD contribution to the generator cancels with the It\^o-Stratonovich correction, we express both terms as the divergence of a double bracket of the modified Hamiltonian. Because of the presence of the L\'evy area correction, the double-bracket dissipation term in \eqref{dissipativeLP} must be changed from $\vartheta \operatorname{ad}_{\frac{\partial C}{\partial \mu}}^{*}\left[\frac{\partial C}{\partial \mu}, \frac{\partial h}{\partial \mu}\right]^{\flat}$ to $\vartheta \operatorname{ad}_{\frac{\partial C}{\partial \mu}}^{*}\left[\frac{\partial C}{\partial \mu}, \frac{\partial \widetilde{h}}{\partial \mu}\right]^{\flat}$. It is with this suitable adjustment of the dissipation to account for the modified Hamiltonian that the same cancellation method can be employed as in \cite{Arnaudon2018}.

The antisymmetry and chain rule properties of the Poisson bracket cancel all of the drift term contributions $\{\mathbb{P}, h + \frac{1}{2}\mathsf{s}^{ij}\{\Phi_i,\Phi_j\}\} = \{\mathbb{P},\widetilde{h}\}$ in the stationary Fokker-Planck equation. The double-bracket and It\^o-Stratonovich terms, when written as a divergence take the form:

\begin{equation}\vartheta \di \left(f(\tilde{h}) \operatorname{ad}_ {\frac{\partial C}{\partial \mu}}\left[\frac{\partial C}{\partial \mu},  
\frac{\partial \tilde{h}}{\partial \mu}\right]\right)
\quad\hbox{and}\quad
\frac{\sigma^2}{2}\di \left(f'(\tilde{h}) \operatorname{ad}_ {\frac{\partial C}{\partial \mu}}\left[\frac{\partial C}{\partial \mu}, \frac{\partial \tilde{h}}{\partial \mu}\right]\right). \label{balance}\end{equation}



In writing these expressions, we have used the divergence-free property of $\ad^*_\mu(\cdot )$ and the chain rule $\frac{\partial}{\partial \mu}f(\widetilde{h}) = f'(\widetilde{h})\frac{\partial \widetilde{h}}{\partial \mu}$. The two terms in \eqref{balance} cancel in the generator \eqref{3.4} and thus the measure $\mathbb{P}(\mu)$ is invariant, provided $f(\widetilde{h}(\mu))$ satisfies $\vartheta f(\widetilde{h}(\mu)) = \frac{1}{2}\sigma^2 f'(\widetilde{h}(\mu))$.  If $f$ satisfies this equation, then $\mathbb{P}(\mu)$ is a Gibbs measure,



\begin{equation}\mathbb{P}(\mu) = Z^{-1}\exp(-\frac{2\vartheta }{\sigma^2 }\tilde{h}(\mu))\,.\label{gibbs}\end{equation} 

This calculation has proved that the L\'evy-area corrected Lie-Poisson equations with double-bracket dissipation 
satisfy the following theorem.

\begin{theorem}\label{thm: Gibbs}The Lie-Poisson equation with double-bracket dissipation and L\'evy-area correction in the following equation 

\begin{equation}\dn \mu+\operatorname{ad}_{(\frac{\partial h}{\partial \mu}\dn t - \mathsf{s}^{ij}[\xi_i, \xi_j] \dn t - \xi \circ \dn W) }^{*} \mu +\vartheta \operatorname{ad}_{\frac{\partial C}{\partial \mu}}^{*}\left[\frac{\partial C}{\partial \mu}, \frac{\partial h}{\partial \mu} - \frac{1}{2}\mathsf{s}^{ij} \left[ \xi_i, \xi_j \right] \right]^{\flat} \mathrm{~d} t = 0\,,
\label{DBDlevy}\end{equation}

admits the following Gibbs measure,

\begin{equation}\mathbb{P}(\mu) = Z^{-1}\exp\left(-\frac{2\vartheta }{\sigma^2 }
\big(h -\big\langle  \mu\,,\, \frac{1}{2}\mathsf{s}^{ij}{[\xi_i, \xi_j]} \big\rangle \big)\right)
\,.
\label{gibbs}
\end{equation} 

\end{theorem}


\section{Example: Stochastic Rigid Body Dynamics}\label{sec:4}

In this section we demonstrate the effects predicted in Theorem \ref{maintheorem} for the specific example of the stochastic rigid body equations. This is a finite dimensional example with a phase space that can be embedded in $\mathbb{R}^3$ and thus is one of few easy to visualise examples available to us. In addition, the stochastic and deterministic rigid body equations are important examples for modelling applications \cite{Casolo2018} \cite{Schomaker1968}. The rigid body is often used as a key example in much of geometric mechanics literature for its rich physical behaviour and ability to draw {analogies to the} infinite dimensional Euler-Poincar\'e equations for fluid dynamics \cite{Marsden1999}. The importance of rigid body dynamics in geometric mechanics warrants a study of the effects of the L\'evy area with techniques from {theory of }rough paths. 

The rigid body equations are defined on the dual of the Lie algebra $\mathfrak{so}(3)$ of the  Lie group $G = SO(3)$ and are obtained from the Lagrangian $l(\Omega) = \frac{1}{2}\Omega \mathbb{I} \Omega =: \frac{1}{2}\Omega \cdot \Pi$, where $\mathbb{I}$ is a fixed diagonal matrix. The Lie bracket needed to write down the Lie-Poisson equations and determine the L\'evy area correction is identified as the cross produce of two vectors in $\mathbb{R}^3$. We include double bracket dissipation, which appears as a vector triple product in the equations. The dissipation can be obtained from the Lagrange-d'Alembert variational principle defined in Section \ref{2.2} with 
\[F_0 = \vartheta \ad^*_{\left({\ad^*_{\Omega {\color{black}-} \frac{1}{2}\mathsf{s}^{ij}[\xi_i, \xi_j]}\Pi}\right)^{\sharp}} \Pi\,,\] 
here $\sharp$ denotes the isomorphism between $\mathfrak{g}^*$ and $\mathfrak{g}$ with inverse map $\flat$.  The L\'evy area term will appear both in the angular velocity and the double bracket dissipation. Consequently, our modification of the stochastic rigid body takes the form,

\begin{equation} 
\dn \Pi {\color{black}-} \Pi \times \Big(\Omega {\color{black}-} \frac{1}{2}\mathsf{s}^{ij} \left(\xi_i \times \xi_j \right)\Big)  \dn t {\color{black}-} \vartheta \Pi \times \left(\Pi \times \left(\Omega {\color{black}-} \frac{1}{2}\mathsf{s}^{ij}\xi_i \times \xi_j \right)\right) \dn t {\color{black}-}  \sum_{i}\Pi \times \xi_i 
\circ \dn W^i = 0.
\label{srblevy}
\end{equation}

We write our equations with Stratonovich Brownian motion as the driver, with the L\'evy area correction appearing as an additional {\color{black}deterministic torque}. The form \eqref{srblevy} will be more convenient to work with as we shall choose known values of $\mathsf{s}$ to avoid having to generate realisations of $\circ\dn \widetilde{W}$ by computer rather than the familiar choice of Brownian motion.   

Although equation \eqref{srblevy} can be mathematically understood in the SDE context without the use of rough paths (since we will always be using a continuous time Markov process to discuss invariant measures) it may be useful to consider \eqref{srblevy} as specifying a rough path that is also a stochastic process, meaning Stratonovich Brownian motion $\mb{B} = (B, \mathbb{B}^\text{Strat})$. The use of the rough paths formalism allows a pathwise interpretation of a solution, which is what one should consider when examining numerical solutions. 

For isotropic noise $\xi_i = \sigma \mathbf{\hat{e}}_i$ the invariant measure as derived in the previous section will be 

\begin{equation}\widetilde{\mathbb{P}}_\infty = \frac{1}{Z}\exp\left(-\frac{\vartheta}{\sigma^2}\Pi \cdot \mathbb{I}^{-1}\Pi {\color{black}+} \vartheta \mathsf{s}^{ij} \Pi \cdot \left(\mathbf{\hat{e}}_i \times \mathbf{\hat{e}}_j\right) \right).\label{measure}\end{equation}

We solve \eqref{srblevy} to obtain the values of $\Pi \in \mathbb{R}^3$ that solve dissipative Lie-Poisson equations with a L\'evy area correction (for details on the numerical methods used see \ref{sec:methods}). The coadjoint motion property ensures that $\Pi$ always lies on the unit sphere which we can project onto a plane for plotting purposes. \newpage

\subsection{Effect of the L\'evy area on the location of equilibrium states}\label{sec:4.1}
 
We now demonstrate the effect of the L\'evy area contribution in the measure in \ref{measure}. 
 
\begin{figure}[H]
    \centering
    \includegraphics[width=0.9\linewidth]{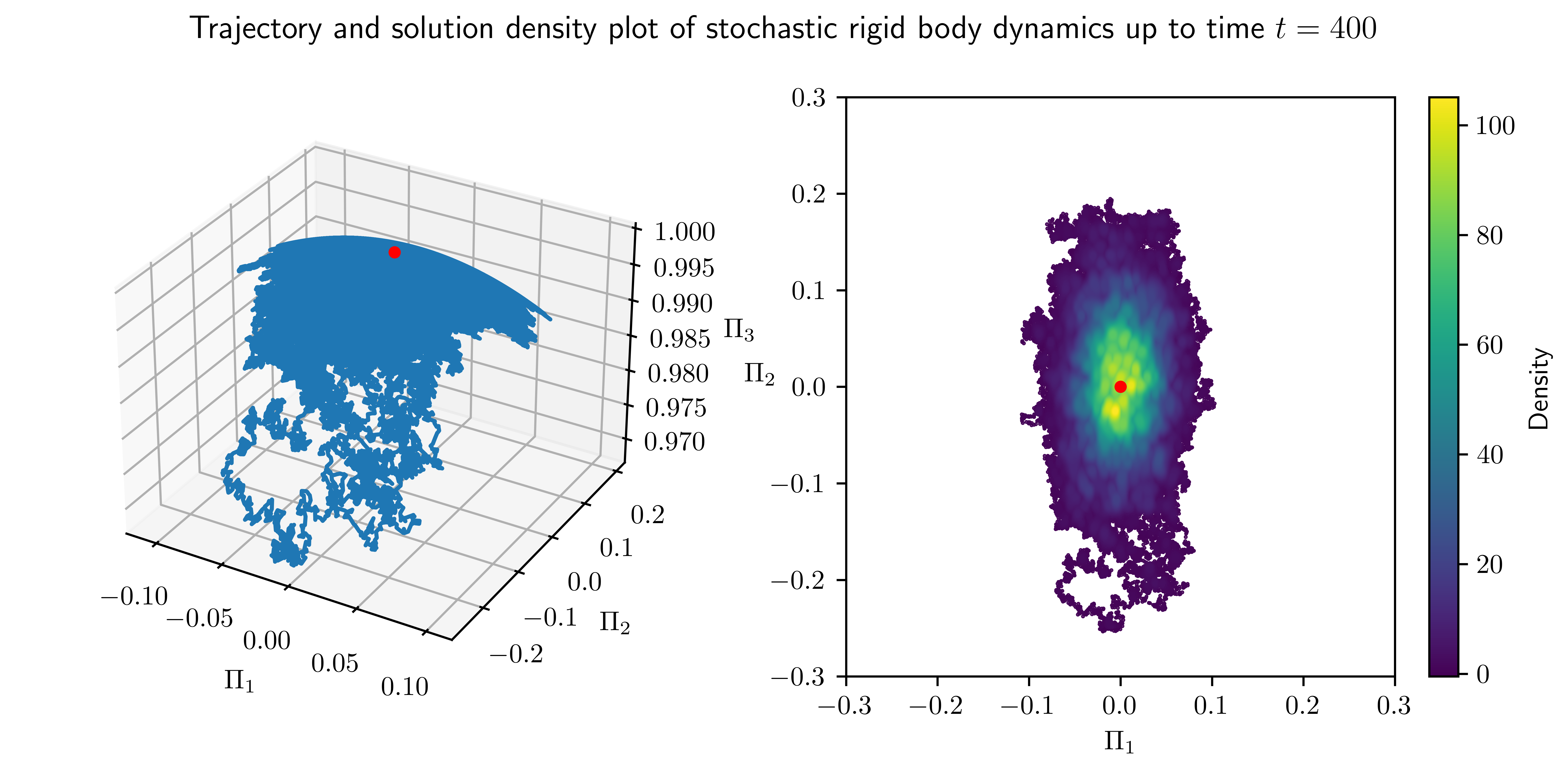}
    \caption{Simulation with $\vartheta = 1, \sigma = 0.05,  \mathbb{I} = \operatorname{diag}(1, 2, 3), \Pi(0) = (0, 0, 1)$, indicated by red dot. The left plot shows the entire trajectory remaining confined to the sphere and moving randomly around the north pole, the right plot demonstrates that the solution is concentrated around the north pole according to the Gibbs measure.}
\label{fig1}\end{figure}

Figure \ref{fig1} shows a side by side solution trajectory and density plot of the same solution projected to $(\Pi_1, \Pi_2)$ plane of the dissipative stochastic rigid body, i.e. \eqref{srblevy} with $\mathsf{s} = 0$. We shall use this as our reference case to compare the effect of the L\'evy area drift term. As mentioned, the geometric nature of the noise chosen ensures that despite random fluctuations in the solution we are always confined onto the unit sphere. Solutions run for a long enough time will converge to behaviour described by the invariant measure, the proportion of time spent will be dictated by the probability density of the measure $\mathbb{P}_\infty$.

Brighter coloured regions denote the solution spending time in a region more often. Depending on the particular initial condition trajectories move to a position of minimum energy, of which there are two, located at the north and south poles $(0,0,\pm 1)$. In the deterministic case without dissipation these are stable centers. The addition of noise causes trajectories to remain near the fixed point in statistical equilibrium around a fixed point described by the Gibbs measure.


\begin{figure}[H]
    \centering
    \includegraphics[width=\linewidth]{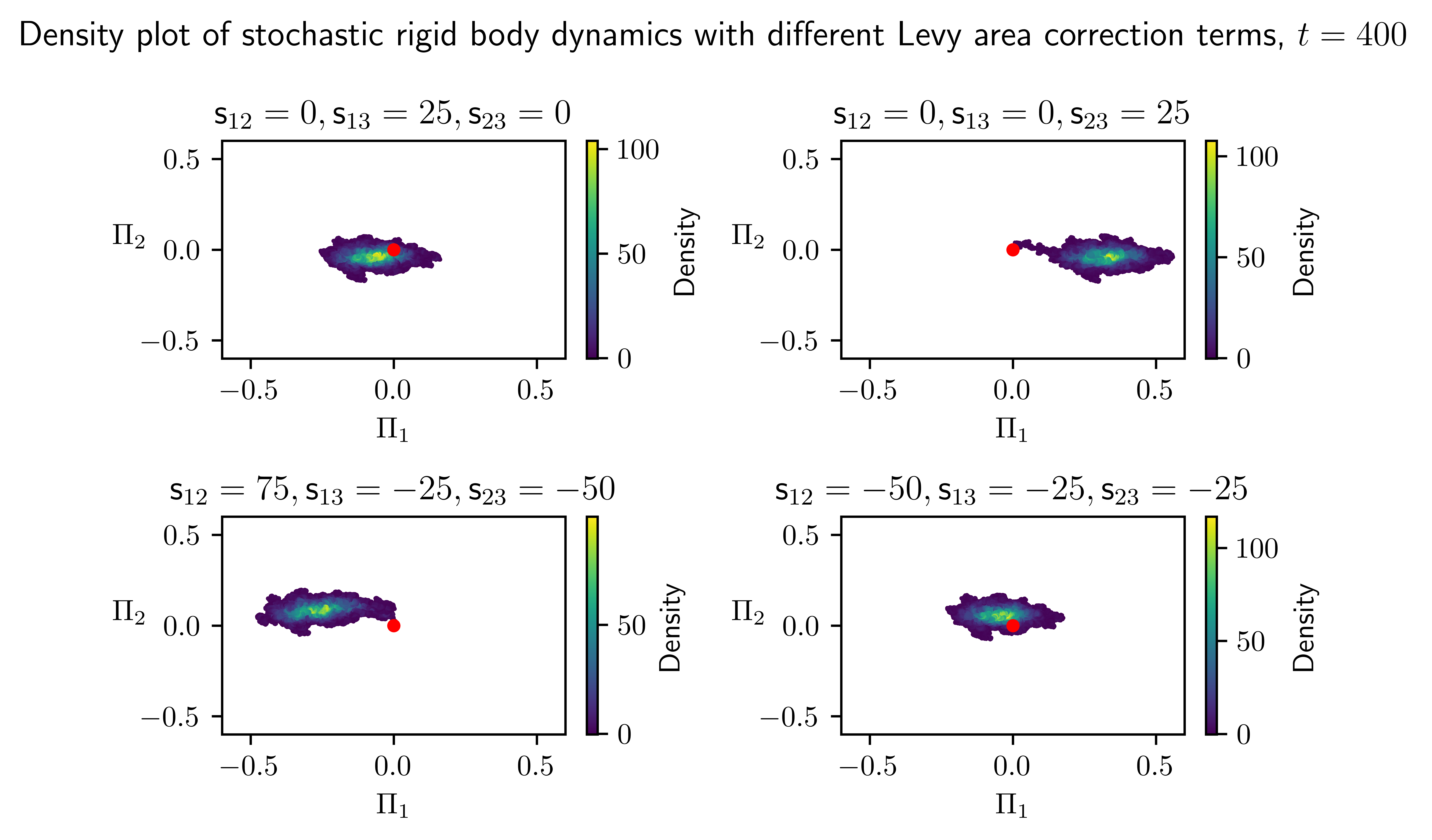}
    \caption{A collection of density plots of various rigid bodies with $\vartheta = 1, \sigma = 0.05,  \mathbb{I} = \operatorname{diag}(1, 2, 3), \Pi(0) = (0, 0, 1)$. Each rigid body has a different choice of L\'evy area perturbation $\mathsf{s}$ and results in a recentring of the Gibbs measure away from the north pole indicated by red dot. The realisation of $B_t(\omega)$ is identical in each subplot to highlight the effects caused by different choices of $\mathsf{s}$.  }
\label{fig2}\end{figure}

Figure \ref{fig2} shows a selection of long time realisations of rigid bodies each with a different choice of non canonical L\'evy areas. We can see that the proportion of time spent in regions is still distributed with a Gibbs measure in \ref{fig1}, but the { additional torque} arising from the L\'evy area correction shifts the mean away from the point north pole at $(0,0)$. The precise location of the recentred means are the subject of the next section. For the two unidirectional L\'evy areas given as the top two plots in Figure \ref{fig2} we can compute these explicitly as $\left(0,3\sigma^2\mathsf{s}^{13}, \sqrt{1 - 9\sigma^4(\mathsf{s}^{13})^2}\right)$ (top left) and $\left(\frac{3}{2}\sigma^2\mathsf{s}^{23}, 0,\sqrt{1 - \frac{9}{4}\sigma^4(\mathsf{s}^{23})^2}\right)$ (top right) as the zeroes of Equation \eqref{detrigidbodylevy} below.


\subsubsection{Deterministic equilibria and Gibbs mean}\label{sec:4.1.1}


We can understand the behaviour found in \ref{sec:4.1} as a consequence of the L\'evy area modifying the deterministic Hamiltonian. Recall that the deterministic rigid body can be expressed as follows, for Casimir $C = \frac{1}{2}|\Pi|^2$:

\begin{equation}
    \frac{d}{dt}\Pi =  - \nabla h \times \nabla C = - \mathbb{I}^{-1}\Pi \times \Pi .
\label{detrigidbody}\end{equation}

The equilibria are determined by solutions of $\Pi \times \mathbb{I}^{-1}\Pi  = 0$. For $\mathbb{I}_1 < \mathbb{I}_2 < \mathbb{I}_3$ rotation around the $\Pi_1, \Pi_3$ axis is Lyapunov stable whilst rotation around the $\Pi_2$ axis is unstable, (see Theorem 15.3.1 \cite{Marsden1999}). Now we consider the contribution of the L\'evy area to the deterministic Hamiltonian $h$, the deterministic dynamics are replaced by

\begin{equation}
    \frac{d}{dt}\Pi = -\nabla \widetilde{h} \times \nabla C = -\left(\mathbb{I}^{-1}\Pi - \frac{1}{2}\mathsf{s}^{ij}[\xi_i, \xi_j] \right) \times \Pi.
\label{detrigidbodylevy}\end{equation}

The equilibrium point $(0,0,1)$ in \eqref{detrigidbody} determines the center of the Gibbs measure when adding noise and dissipation as seen in Figure \ref{fig1}. It is clear that $(0,0,1)$ need not be an equilibrium point for \eqref{detrigidbodylevy}. However, the particular case that $\mathsf{s}^{13} = \mathsf{s}^{23} = 0, \mathsf{s}^{12} \neq 0$ is an example where the stable nodes $(0,0,\pm 1)$ persist in both \eqref{detrigidbody}, \eqref{detrigidbodylevy}. For this choice the L\'evy area contributes only in $\Pi^3$ axis. Since the L\'evy area's contribution vanishes in the derivative, applying the energy Casimir method \cite{Marsden1999} gives the same stability as before. The conclusion from this analysis is that when we modify the L\'evy area of the stochastic rigid body only in the $\mathsf{s}^{12}$ component, the centering of the Gibbs probability measure is unaffected as a consequence of the determinstic equilibria at $(0,0,\pm 1)$ retaining their position and stability. 

\begin{figure}[H]
  \centering
  \begin{minipage}[b]{0.4\textwidth}
    \includegraphics[width=\textwidth]{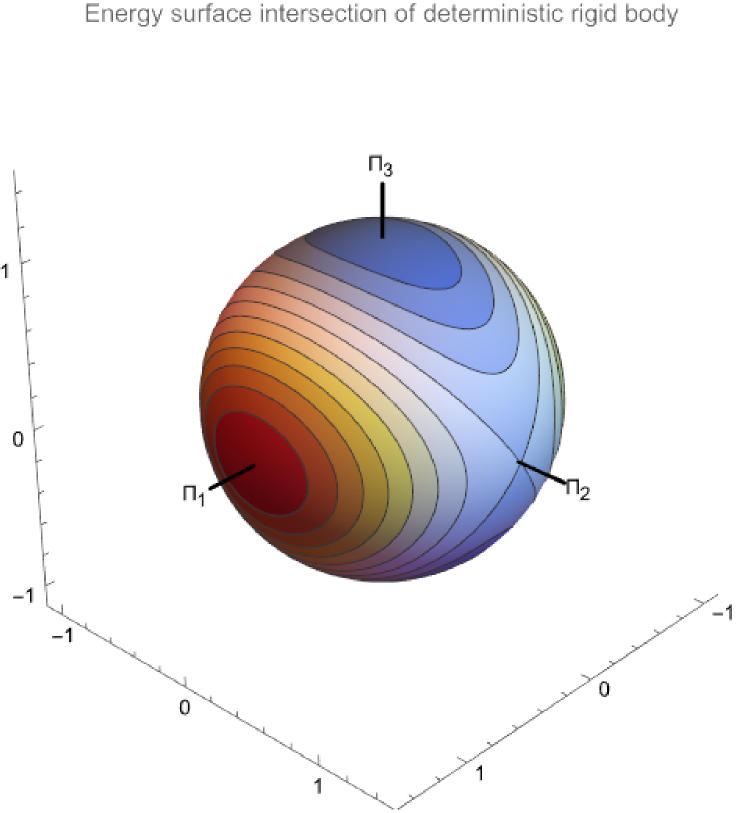}
    \caption{ }
  \label{surf}\end{minipage}
  \hfill
  \begin{minipage}[b]{0.4\textwidth}
    \includegraphics[width=\textwidth]{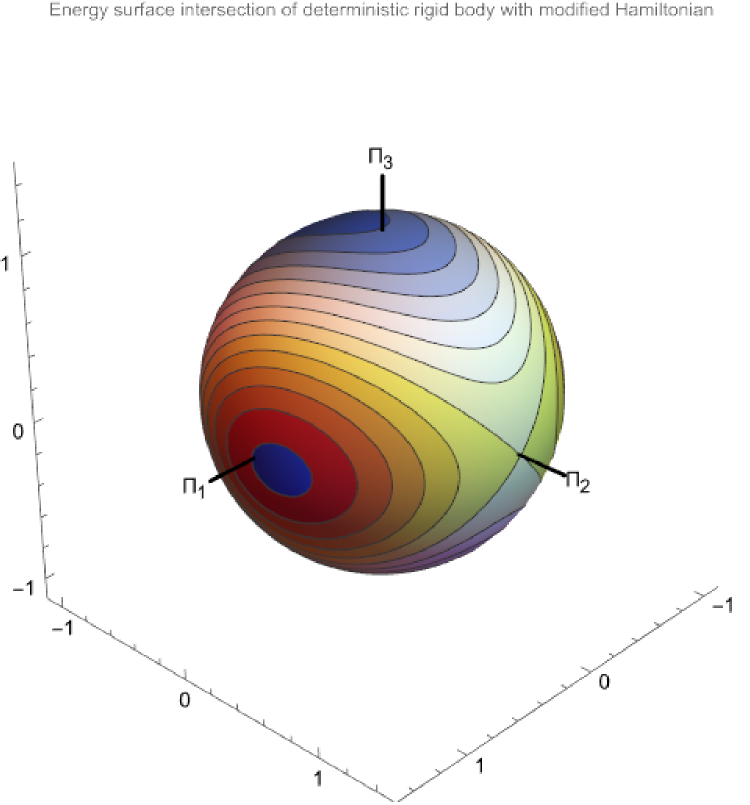}
    \caption{ }
  \label{surf2}\end{minipage}
  \caption*{Figures \ref{surf} and \ref{surf2} show the intersection of the surface defined by $|\Pi|^2 = const$ with various level sets of the Hamiltonian $h(\Pi) = \frac{1}{2}\Pi \cdot \mathbb{I}^{-1} \Pi$ (shown left) and the modified Hamiltonian $\widetilde{h}(\Pi) = \frac{1}{2}\Pi \cdot \mathbb{I}^{-1} \Pi + 0.0625\Pi \cdot \mathbf{\hat{e}}_2$ (shown right). This modification of the Hamiltonian in Figure \ref{surf2} corresponds to an isotropic noise $\xi_i = \sigma \mathbf{\hat{e}}_i$ with intensity $\sigma = 0.05$, and a L\'evy correction of $\mathsf{s}^{13} = -\mathsf{s}^{31} = 25$. These coefficients result in $\Pi \cdot \frac{1}{2}\mathsf{s}^{ij}[\xi_i, \xi_j] = \Pi \cdot \big( \frac{1}{2}\sigma^2 \mathsf{s}^{13} \mathbf{\hat{e}}_1 \times \mathbf{\hat{e}}_3 + \frac{1}{2}\sigma^2 \mathsf{s}^{31} \mathbf{\hat{e}}_3 \times \mathbf{\hat{e}}_1 \big)= -0.0625 \Pi \cdot \mathbf{\hat{e}}_2$ cf. equation \eqref{modifiedhamiltonian}. A stochastic rigid body with noise possessing these parameter values is simulated in the top left of Figure \ref{fig2}.} 
\end{figure}

For general L\'evy area corrections outside the case described above, the equilibrium solution at $(0,0,1)$ will be moved, among others. Figures \ref{surf} and \ref{surf2} illustrate the movement of these equilibria in the rigid body's phase portrait for one simple choice of L\'evy area correction. The rigid body phase portrait is formed by the intersections of the Hamiltonian with the angular momentum sphere $|\Pi|^2=const$. The shift of the Hamiltonian along the $\Pi_2$ axis due to the L\'evy area in \ref{surf2} breaks the reflection symmetry of the unperturbed phase portrait. 

The key differences between Figures \ref{surf} and \ref{surf2} are that instead of four heteroclinic orbits connecting the two saddle points on opposite poles of the angular momentum sphere along the $\Pi_2$ axis, the L\'evy area shift produces two pairs of homoclinic orbits forming figures of eight on the sphere centred at each pole of the $\Pi_2$ axis, and the relative orientation of the two figures of eight is rotated by $\pi/2$. (See appendix Figure \ref{surf3} to see a view of the opposing side of the sphere.) The stable centers that once were located at $(\pm 1,0,0)$ and $(0,0,\pm 1)$ have all been moved forward in the $\Pi_2$ direction, and are now enclosed within the newly created homoclinic orbits. Although in the context of this paper we do not smoothly vary $\mathsf{s}$ (as this would correspond to continuously changing our interpretation of the stochastic integral), the effect of $\mathsf{s}$ on the deterministic dynamics in Figures \ref{surf} and \ref{surf2} is essentially the creation of a global bifurcation.

\subsection{Rigid body dynamics driven by coloured noise}

In section \ref{sec:4.1} we have shown that the deterministic torque caused by the L\'evy area produces similar behaviour to that of the dissipative stochastic rigid body (a Gibbs invariant measure) with a shift in the centre of the Gibbs probability distribution. The simulation of Eq. \eqref{srblevy} can be thought of as a stochastic RDE with rough driver $(B_t - B_s, \mathbb{B}^{\text{Strat}}_{s,t} + \mathsf{s}(t-s))$ which is then translated it back to an SDE with additional drift using Proposition \ref{rdeproposition}. As shown in figures \ref{surf} and \ref{surf2} the aforementioned drift terms produce a non trivial effect on the dynamics.

Having established the potential effects of the L\'evy area correction one is interested in how specific modelling choices may lead to the appearance of such correction. Particularly, we are interested in showing how limits of the kind seen in homogenisation theory \cite{Pavliotis2008} and use of coloured Ornstein-Uhlenbeck type noise, or physical noise processes \cite{Clark1966}, lead to the situation of $\mathsf{s} \neq 0$.  

The class of physical noise processes described in \cite{Clark1966} encompass those of piecewise continuous approximations seen in \cite{Ikeda1981}, as well as the physical Brownian motion in \cite{Friz_2015}. The scaling of physical Brownian motion coincides with the square root of the covariance matrix factor for Ornstein-Uhlenbeck noise coloured noise seen in homogenisation theory. An assumption within this class of physical noise processes is that the iterated integrals converge to the iterated integrals of Brownian motion acted on the left by a constant matrix $B$ and shifted by a matrix $A(t-s)$, known as the characteristic matrix. In the modern formulation of rough paths this is saying physical noise processes converge to the signature of the rough path lift of $BW_t$ with a shift in the area enhancement. The symmetric part of the characteristic matrix in \cite{Clark1966} is precisely the L\'evy area correction, that is $\operatorname{Alt}(A) = \mathsf{s}$. The case of symmetric $A$ is discussed in \cite{Clark1966} within the context of recovering the Wong-Zakai theorem \cite{Wong1965a, Wong1965b}.

In this paper we shall only consider a particular case of such physical noise processes that lift to a geometric rough path and allow us to exactly calculate quantities determining the L\'evy area, but we remark that a wide class of stochastic processes considered in \cite{Clark1966} can be considered that will lead to a modification to the signature of the rough path and produce interesting dynamical effects driven by noise.


The manner of how noise with $\mathsf{s} \neq 0$ can influence dynamics is clearly illustrated by considering the following planar equation with an OU type coloured noise approximation.
\begin{equation}
\begin{aligned}
\dn x &= Qx\dd t +  P(x) 
 \frac{y}{\varepsilon}\dn t \\
\dn y &= -\frac{1}{\varepsilon^2}(\alpha I + \gamma J)y\dn t +  \frac{\sqrt{\delta}}{\varepsilon} \circ \dn B \, ,
 \end{aligned}\label{2dexample}\end{equation}

for $Q = \Big(\begin{smallmatrix}
  \lambda & 0\\
  0 & \mu
\end{smallmatrix}\Big)$, $P(x) = \Big(\begin{smallmatrix}
  0 & x_2\\
  x_1 & 0
\end{smallmatrix}\Big)$ and $J = \Big(\begin{smallmatrix}
  0 & 1\\
  -1 & 0
\end{smallmatrix}\Big)$, with $\alpha,\gamma, \delta$ positive constants and $\lambda, \mu$ chosen depending on whether stable or unstable behaviour is desired. Equation \eqref{2dexample} is a special case of eq. 5.16 in \cite{Pavliotis2014}[Sec. 5.1]. Referring to the homogenisation theory interpretation of this equation, it is known in \cite{Pavliotis2014} that the $\varepsilon \rightarrow 0$ limit converges to a new system with additional drift given by \footnote{Here the divergence of a matrix valued function is defined $(\nabla \cdot M)_i := \partial_j M_{ij}$} $b(x) =\frac{\delta \gamma}{2 \alpha (\alpha^2 + \gamma^2)}\Big(\nabla \cdot\big(P(x) J P(x)^T\big)-P(x) J^T \nabla \cdot P(x)^T\Big)$, furthermore, this $b(x)$ can be shown to be equal to precisely the Lie bracket $[P^1(x), P^2(x)] = \Big(\begin{smallmatrix}
  x_1 & 0\\
  0 & -x_2
\end{smallmatrix}\Big) =: Sx$ times the constant $\frac{\delta \gamma}{2 \alpha (\alpha^2 + \gamma^2)}$. This constant is the L\'evy area $\frac{1}{2}\mathsf{s}^{12} - \frac{1}{2}\mathsf{s}^{21}$ when we interpret the $\varepsilon \rightarrow 0$ limit of \eqref{2dexample} in the rough paths sense of \cite{Friz_2015}.

We can express the limiting system as:
$$\dn X = \left( Q + \mathsf{s}^{12}S\right)X \dn t + \sqrt{\frac{\delta}{\alpha^2 + \gamma^2 }}P(X) \circ \dn W .$$

Because of the form $S$ takes, we see that the spectrum of the original linear system has been modified by the L\'evy area. For particular choices of $\lambda, \mu$ and $\mathsf{s}^{12}$ it is evident that we can engineer circumstances where the approximation of a system with say, a stable node results in limiting equations without this property.

With the example of \eqref{2dexample} in mind we consider the dissipative stochastic rigid body, with no L\'evy area terms. We replace the noise term $\dn W$ by an approximation of the standard white noise and we expect the possibility that a L\'evy area correction will appear with a noticeable effect on solution behaviour. Our choice of approximation of white noise is a Gaussian stationary stochastic process $y^\varepsilon_t$ that formally satisfies $\lim\limits_{\varepsilon \rightarrow 0}\mathbb{E}[\frac{y_t}{\varepsilon} \otimes \frac{y_s}{\varepsilon}] \propto I\delta(t-s)$ and that its time integral converges to a Brownian motion.

Following from the above example, we take Ornstein-Uhlenbeck noise solving the following stochastic differential equation 

\begin{equation}
 \dd y_t = -\frac{1}{\varepsilon^2}Ay_t\dd t + \frac{1}{\varepsilon}D \circ\dd B_t , \label{coloured}\end{equation}

for a given $A, D$. (See the appendix for a proof that the integral of this process does indeed converge to Brownian motion and conditions on $A,D$.)

It is shown in \cite{Clark1966, Friz_2015, Hairer2020} that the anti symmetric part of $A$ in this type of physical Brownian motion/coloured noise will cause $\mathsf{s} \neq 0$. It so happens that this form of noise provides an elementary example of a stochastic process that converges to a rough path with a non-canonical L\'evy area, and that this L\'evy area correction can be explicitly calculated for certain cases. In the appendix we show that for a coloured noise given in the form of \eqref{coloured} that

\begin{equation}\mathsf{s} = \Sigma (A^{-1})^T - A^{-1} \Sigma.\label{levycoloured}\end{equation}

Where $\Sigma$ is the variance-covariance matrix of the stationary Gaussian distribution of the coloured noise rescaled to be independent of $\varepsilon$. The matrix $\Sigma$ can be calculated as in \cite{Clark1966, Friz_2015, Hairer2020} to be:

\begin{equation}\Sigma = \int_{-\infty}^0\exp(\rho A)DD^T\exp(\rho A^T)d\rho.\label{covariance}\end{equation}

For the case of the stochastic rigid body in $\mathbb{R}^3$, we will take $D = I_{3\times 3}, A = kI_{3 \times 3} + \Lambda$ where $\Lambda = \operatorname{skew}(\alpha,\beta,\gamma)$ is some arbitirary $3 \times 3$ matrix satisfying $\Lambda^T = -\Lambda$ and some constant $k$. The example we shall use is an adaptation of a very simple planar system in \cite{Pavliotis2008} that is amenable to calculation. Note that $A, A^T$ commute in this case and the antisymmetric portions in the integral cancel. Our choice of example allows us to calculate the integral \eqref{covariance} explicitly as $I/2k$, in this case we have $\mathsf{s} = -\frac{1}{k}\operatorname{Alt}(A^{-1})$. 

To demonstrate the L\'evy area's effect on the approximation problem of the ordinary stochastic rigid body, we solve the equation:

\begin{equation} \dn \Pi {\color{black}-} \Pi \times \Omega \dn t {\color{black}-} \vartheta \Pi \times \left(\Pi \times \Omega \right) \dn t {\color{black}-}  \sum_{i}\Pi \times \xi_i \frac{y^i}{\varepsilon}\dn t = 0,\label{srbcoloured}\end{equation}
\begin{figure}[H]
    \centering
    \includegraphics[width=\linewidth]{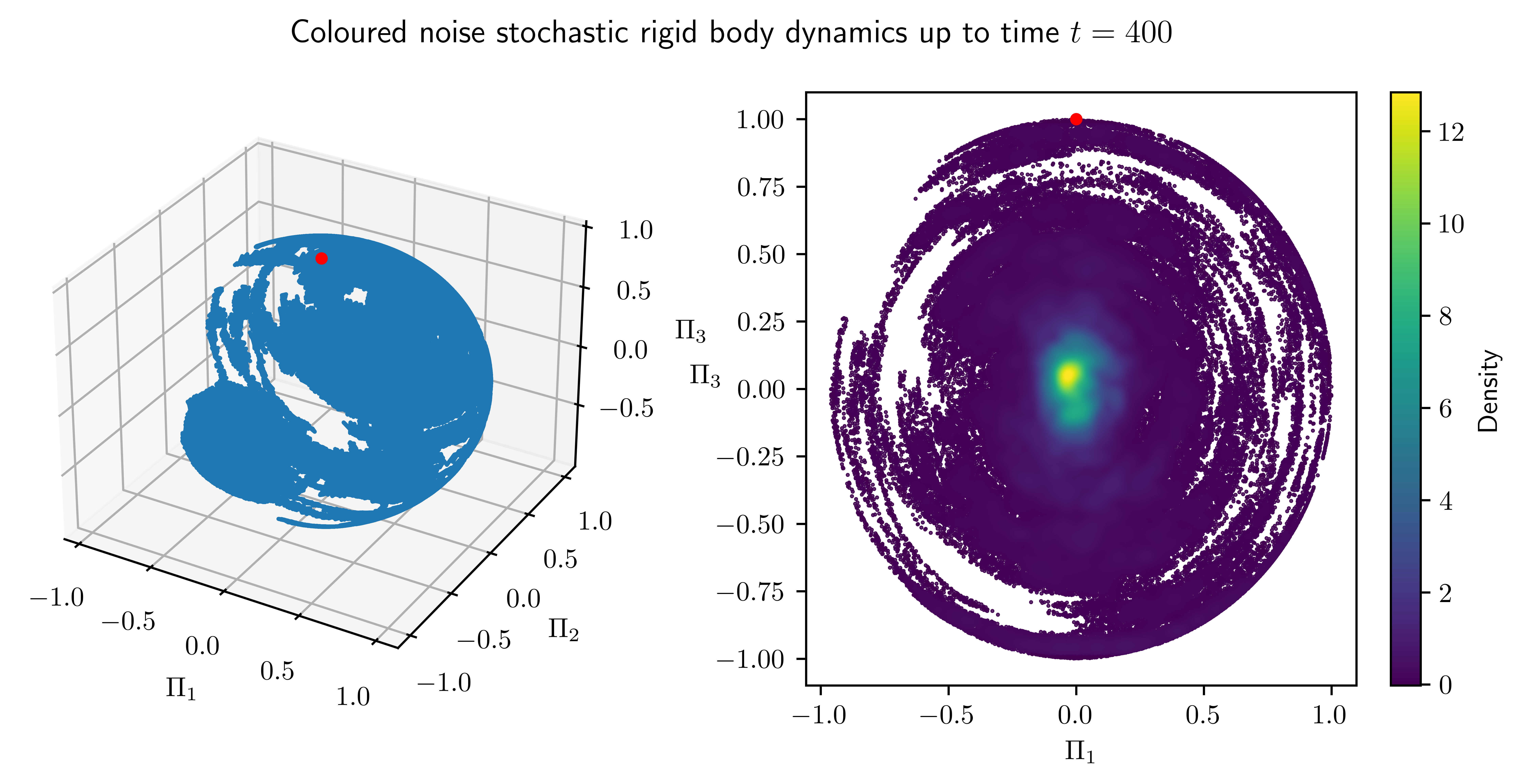}
    \caption{Simulation with $\vartheta = 1, \sigma = 0.05,  \mathbb{I} = \operatorname{diag}(1, 2, 3), \Pi(0) = (0, 0, 1)$. The solution starts at the north pole marked by the red dot, and is immediately forced away. As can be seen the solution eventually spirals towards $(0,1,0)$ and is distributed with a near Gibbs invariant measure, centered in a dramatically different position.}
\label{fig4}\end{figure}

Figure \ref{fig4} shows a coloured noise approximation of the stochastic rigid-body \eqref{srbcoloured} in the $(\Pi_1, \Pi_3)$ plane for $\varepsilon = 0.005, \Lambda = \operatorname{skew}(0,0,1)$ and $k = 0.05$. The choices of $k, \gamma$ correspond to a value of $\mathsf{s}^{13} \approx 19.95$ in the coloured noise limit with $\mathsf{s}^{12} = \mathsf{s}^{23} = 0$, but we stress that the equations write down and solve have \emph{no additional drift}, the L\'evy area manifests itself as a consequence of failing to converge to the correct signature of the rough path.

The resulting behaviour for the coloured noise approximation is drastically different than what is expected in \ref{fig1}, now the solution has been forced away from the north pole completely and remains near $(0,1,0)$, which along with $(0,-1,0)$ are the only persisting fixed points from the original rigid body for $\mathsf{s}^{13} \neq 0$. 

As the coloured noise approximation converges to $A^{-1}D\dd B_t$ instead of $\dd B_t$ (i.e. correlations have been introduced) the limiting system is driven by non-isotropic noise. Although we cannot deduce an analytic expression for the invariant measure exactly of the form \eqref{measure} without the isotropic assumption we can see the behaviour still approximates a Gibbs distribution, agreeing with the non isotropic example in \cite{Arnaudon2018} [Fig. 1]. The equation \eqref{detrigidbodylevy} can be used to determine the location of deterministic equilibria but because of the directional bias caused by \[
A^{-1}D = \begin{pmatrix}
  k/{(k^2+1)} & 0 & -1/{(k^2+1)}\\
  0 & 1/k & 0\\
  1/{(k^2+1)} & 0 & k/{(k^2+1)}
\end{pmatrix}
\approx \begin{pmatrix}
  0.05 & 0 & -1\\
  0 & 20 & 0\\
  1 & 0 & 0.05
\end{pmatrix}
\] we cannot expect solutions to follow the movement of the nearest fixed point to $(0,0,1)$ as they did in Figure \ref{fig2}. We can verify that the coloured noise approximation indeed converges to scaled Browmian motion with the correct choice of L\'evy area in the following figure.

\begin{figure}[H]
    \centering
    \includegraphics[width=\linewidth]{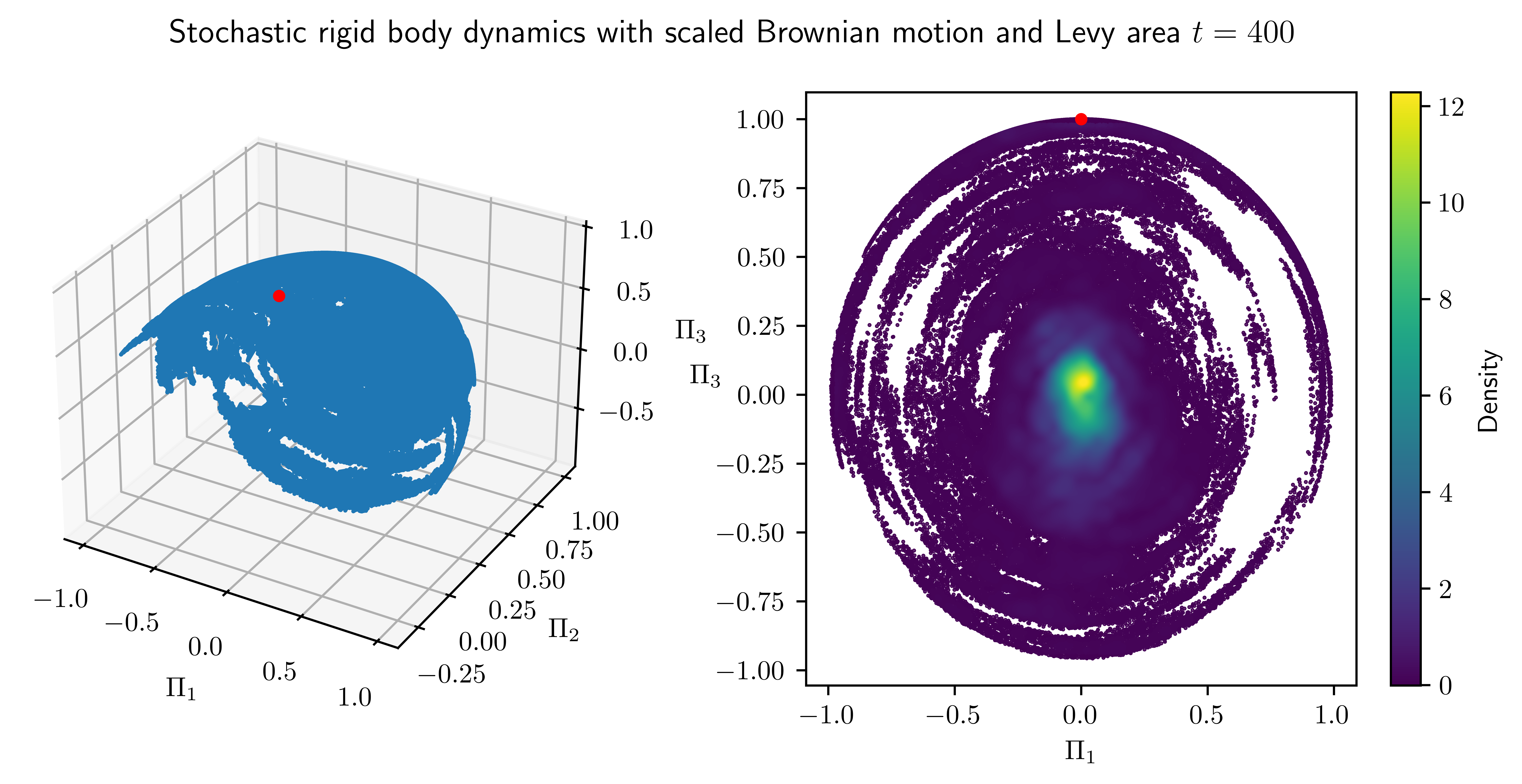}
    \caption{Simulation with $\vartheta = 1, \sigma = 0.05,  \mathbb{I} = \operatorname{diag}(1, 2, 3), \Pi(0) = (0, 0, 1)$ and a L\'evy area of $\mathsf{s}^{13} \approx {19.95}$. We have also scaled the Brownian noise $(\dn B^1, \dn B^2, \dn B^3)$ by the matrix $A^{-1}D$. The simulated solution behaviour nearly matches that of Figure \ref{fig4}, where the solution spirals away from the north pole and settles around $(0,1,0)$.}
\label{fig5}\end{figure}

Figure \ref{fig5} shows the associated white noise rigid body with a L\'evy area correction, as well as a scaling matrix of $A^{-1}D$ applied to the Brownian motion. As can be seen the behaviour aligns with the behaviour seen in Figure \ref{fig4}, which agrees with the prediction that the coloured noise approximation is converging to equation \eqref{srblevy}. The movement to the new region can be explained using the analysis in the previous section \ref{sec:4.1.1}, the equilibrium at $(0,0,1)$ no longer exists, while the unstable equilibrium at $(0,1,0)$ persists. The L\'evy area does not change the saddle point behaviour at $(0,1,0)$ and so we observe a slow spiraling towards the $\Pi_2$ axis, this is further directed by stronger noise component in the $\Pi_2$ direction of $A^{-1}D \dn B_t$ originating from asymmetries in the matrix $A$ which generate a L\'evy area correction and scaling. 

The figures in this section are demonstrations of what is often referred to as the Wong-Zakai anomaly. Equations of the form \eqref{srbcoloured} approximate the behaviour of \eqref{srblevy}, rather than the expected stochastic rigid body solution in Figure \ref{fig1} with no additional torque terms. As predicted in the results of Section \ref{sec:3} we have disagreement in long time behaviour via changes in the Gibbs measure. Namely, convergence to the north pole which is expected for the stochastic rigid body with dissipation does not always occur due to the additional torque of the L\'evy area. We instead observe convergence to a separate region of the sphere depending on the nature of the noise we select. 

\section{Summary and concluding remarks}\label{sec:Conclusion}

In this paper we have studied the effect of the L\'{e}vy area correction on the qualitative and quantitative properties of dynamics derived from variational principles on GRP. We began with a modification of the rough Hamilton-Pontryagin variational principle for SALT that includes the additive contributions of the L\'evy area to the angular velocity of the rigid body. In Section \ref{sec:2} we found that the L\'evy area correction term preserves the coadjoint properties of the Lie-Poisson system, although its modification of the angular velocity introduces an additional deterministic torque in the evolution equation for the angular momentum of the rigid body. 

In Section \ref{sec:3} we discussed reasonable conditions for the law of solutions to be equivalent under a change in L\'evy area and demonstrated how the preserved Lie-algebraic properties also explain why the long time behaviour leads to a constant invariant measure on the initial coadjoint orbit. Next, we introduced a double-bracket type of dissipation that preserves the coadjoint orbits, while sending the solutions toward a minimal energy state on the level set of the Casimir surface containing the initial conditions. When the double-bracket dissipation was applied, Theorem \ref{thm: Gibbs} showed that the stochastic Lie-Poisson solutions at long times tended toward a Gibbs distribution. However, the properties of the emerging Gibbs distribution were also affected by the L\'evy area. Specifically, Theorem \ref{thm: Gibbs} showed that inclusion of the L\'evy area preserved the form of the Gibbs measure, while it modified the energy in the argument of the Gibbs distribution exponential. The results of Theorem \ref{thm: Gibbs} were exemplified for GRP rigid body dynamics with double-bracket type of dissipation. We also simulated the invariant measure for rigid body motion whose angular velocity was augmented by coloured noise and demonstrated that the L\'evy area can be understood from the theory of rough paths as a Wong-Zakai anomaly.

There are many natural research directions to be developed further in rough path geometric mechanics. This paper has considered the Lie-Poisson dynamics on geometric rough paths that is generated by a finite dimensional Lie group acting on the tangent space of a configuration manifold. Here, the rigid body example reveals the effects of the L\'evy area correction on coadjoint motion in the presence of double bracket dissipation. 
In the light of Theorem \ref{thm: Gibbs}, which is posed abstractly for the Lie-Poisson equations, one can imagine that the conclusions for the rigid body example derived here extend quite analogously in other finite dimensional examples of coadjoint motion on geometric rough paths. As a matter of fact, the possibilities for its extension are much better, since the contribution of the L\'evy area will always be simply a shift of the original unperturbed deterministic Hamiltonian of the same form as seen in the Gibbs distribution in \eqref{gibbs}. Symbolically the expressions defined throughout section \ref{sec:2} are well defined for equations posed on general, possibly infinite dimensional, Lie groups such as the group of diffeomorphisms. The extension to general Lie group equations will require additional work to compensate for loss of arguments involving structure constants and the existence of a bi-invariant Killing form.  

Since the present study has only involved kinetic energy, another natural extension would be to incorporate potential energy into the Lie-Poisson formulation of Hamiltonian dynamics on geometric rough paths. This extension may be accomplished by promoting the Lie group action on the configuration manifold to semidirect-product action, $S = G \ltimes V$, where $V$ is a vector space upon which $G$ acts via linear maps and whose multiplication rule is given by:

$$\left(g_{1}, v_{1}\right) \cdot \left(g_{2}, v_{2}\right):=\left(g_{1} g_{2}, v_{1}+g_{1} v_{2}\right) .$$

{\color{black}The effects of potential energy on the dynamics are encoded in the diamond operator $\diamond : V \times V^* \rightarrow \mathfrak{g}^*$, which represents the dual semidirect-product Lie algebra action defined as} 
\[
\langle v \diamond a, \xi \rangle_{\mathfrak{g}^*\times\mathfrak{g}} = \langle v, -\,\xi a \rangle_{V^*\times V}\,,
\]
with $a\in V$, $v\in V^*$ and $\xi a$ denoting the Lie algebra action for $\xi\in\mathfrak{g}$ acting on $a\in V$. In the stochastic case, this Lie algebra action on $V$ eventually leads to stochastic Lie-Poisson equations {\color{black}expressed on the dual of} a semidirect-product Lie algebra as,

$$\begin{aligned}
\dn \mu &= - \mathrm{ad}_{\frac{\delta h}{\delta \mu}}^{*} \mu \dd t - \mathrm{ad}_{\xi}^{*} \mu \circ \dn B + \frac{\delta h}{\delta a} \diamond a \dd t\,, \\
\dn a &= \frac{\delta h}{\delta \mu} a \dd t + \xi_i a \circ \dn B^i\,,
 \end{aligned}$$

We can then consider generalisations to rough paths as in \cite{Leahy2020}. The $\mu$ dynamics pick up the same correction term as in the proof of Proposition \ref{rdeproposition}. For the advected quanitity $a$, proceeding as in \ref{rdeproposition}, we calculate the Gubinelli derivative of $(\xi a)^\prime_{ij} = \xi_i (\xi_j a)$. Under the assumption that $\mathfrak{g}$ is a matrix Lie algebra we are able to antisymmetrize and write the L\'evy area correction as $\frac{1}{2}\mathsf{s}^{ij}[\xi_i, \xi_j] a$. What is remarkable is that in both of the resulting equations we have the same modified Hamiltonian $\widetilde{h} = h -\big\langle  \mu\,,\, \frac12\mathsf{s}^{ij}{[\xi_i, \xi_j]} \big\rangle$ appearing in $\ad^*$ for the $\mu$ evolution and acting on the advected quantity $a$, with no incompatibility when deriving the corrected $\widetilde{h}$ from either equation.  The heavy top equations are a natural extension of the rigid body to investigate. From the formal equations derived above with the deterministic Hamiltonian $h(\Pi, \gamma) = \frac{1}{2}\Pi \cdot \mathbb{I}^{-1}\Pi + \chi \cdot \gamma$ the system takes the form:

\begin{equation}\begin{aligned}
    \dn \Pi &= \Pi \times \Big(\Omega {\color{black}-} \frac{1}{2}\mathsf{s}^{ij} \left(\xi_i \times \xi_j \right)\Big)  \dn t  +  \sum_{i}\Pi \times \xi_i  
\circ \dn W^i + \chi \times \gamma \dd t \\
\dn \gamma &= \gamma \times \Big( \Omega {\color{black}-} \frac{1}{2}\mathsf{s}^{ij} \left(\xi_i \times \xi_j \right)\Big)\dn t + \sum_i \gamma \times \xi_i \circ \dn W^i \,.
\end{aligned}\label{heavytoplevy}\end{equation}

Here the advected quantity $\gamma \in V = \mathbb{R}^3$ known as the gravity vector models how gravity is oriented with respect to the top's position and the constant vector $\chi \in \mathbb{R}^3$ points in the direction of the center of mass of the body to a fixed point of support or "pivot". It can be deduced immediately from taking dot products with $\gamma, \Pi$ and adding the two equations that the Casimirs $|\Gamma|^2, \Gamma \cdot \Pi$ continue to be conserved quantities for \eqref{heavytoplevy} and that the L\'evy area does not impact this. 

Other extensions and geometric variants of the SALT approach to the addition of GRP transport examined in this paper also exist. For example, the physical and statistical effects of the L\'evy area on the  Lagrangian averaged (LA) SALT \cite{Drivas2020} model, which produces Euler-Poincaré and Lie-Poisson equations, will have a change in expected velocity and a modification to the Hamiltonian as discussed in this paper. The LA SALT approach models the dynamics of expectations, variances and other "climate" features of the respective SALT equations studied here. In particular, the decaying magnitude of expectations and growth of fluctuations from the mean shown in \cite{Drivas2020}[Sec. 4.1] for the LA SALT rigid body are related to its uniform invariant measure in the absence of dissipation. The extension to include dissipation and L\'evy area correction terms for LA SALT will result in nonuniform invariant measure and different fluctuation dynamics. In addition, companion models to SALT with related geometric structures exist. One promising example is SFLT (stochastic \emph{forcing} by Lie transport) \cite{Hu2021}. The corresponding Eulerian averaging (EA) SFLT could also be studied similarly and its results compared with LA SALT.  


Yet another direction of current interest is the relation with deterministic homogenisation \cite{ilya2019, Kelly2017}. In \cite{Cotter2017} it is shown that the equation for stochastic Lagrangian particle dynamics

\begin{equation}\dn q = u(q,t) \dn t + \sum_i^M \xi_i(q) \circ \dn W^i,\label{homog}\end{equation}

can be rigorously derived via a decomposition of the flow map $q$ into the composition of a slow mean field map and a fast map, provided one makes a dynamical systems assumption of being "sufficiently chaotic", as discussed in \cite{Cotter2017},

$$q(X,t) = \bar{q}(X,t) + \zeta_{\frac{t}{\varepsilon}} \circ \bar{q}(X,t).$$

The stochasticity in this system arises from uncertainty in observing or resolving the fast motion. One also decomposes $\partial_{t} \zeta\left(\overline{q}(X, t), \frac{1}{\varepsilon} t\right)=\sum_{i=1}^{M} \lambda_{i}\left(\frac{t}{\varepsilon}\right) \phi_{i}(\overline{q}(X, t))$ into slowly varying eigenfunctions. One asks whether in the limit of widely separated time scales there might exist a type of homogenisation for fluid dynamics that would lead naturally to geometric rough paths. This, of course, is an open problem. 

One approach to this problem of interest to us rests on the possibility that the homogenisation procedure used to derive SALT for fluids could also converge to neither It\^o nor Stratonovich stochastic integrals, i.e. it would produce correction terms. Rough path techniques can  be used to prove a weak invariance principle for $W_n := \frac{1}{\sqrt n}\int_0^{nt} \partial_t \zeta_t(\lambda(s)) \dn s$ and the enhancement $\mathbb{W}_n := \int W_n \otimes \dn W_n$. It is known that this limit can occasionally produce nontrivial L\'evy areas \cite{Gottwald2022}. If one could derive \eqref{homog} to include L\'evy area correction, then this would provide a homogenisation theory viewpoint for equations such as \eqref{velocity}, which in turn would lead to the variational principles defined and used in this paper. 

It is important to note that the stochastic rigid body can \emph{not} fit into this formalism for a finite dimensional example, as the decomposition of diffeomorphisms into the sum of slow and fast components requires an addition operation unavailable when working in $SO(3)$, this is all the more reason to investigate the L\'evy area's effect on infinite dimensional models. When adjusted for right invariant Lagrangians the variational principle \eqref{varprinciple} is well defined for infinite dimensional Lie groups, although the results discussed in Section \ref{sec:3} very much depend on the properties of finite dimensional semisimple Lie algebras. An interesting question could be how the L\'evy area impacts known properties of equations such as wave breaking and peakon formation for the stochastic Camassa-Holm equation, which was examined in \cite{Crisan2018} as a fundamental SPDE model of uncertainty in nonlinear wave dynamics. A caveat here is that while the L\'evy area may produce a leading order deterministic effect on the centre of probability of the invariant measure, the determination of the L\'evy area from observed or simulated data, say, is not sufficient to uniquely determine the individual vectors $A_n(X_t)$ in the definition \eqref{ikedawatanabe} of the geometric rough paths.  

{\color{black}
\subsection*{Acknowledgements}
We thank our colleagues who offered encouraging comments and vital information during this work. These friends and colleagues include D. Crisan, M. Hairer, J.-M. Leahy, Erwin Luesink, W. Pan, X.-M. Li and the Friday afternoon Geometric Mechanics Research Group.  
DH is grateful for partial support from ERC Synergy Grant 856408 - STUOD (Stochastic Transport in Upper Ocean Dynamics ).
The work of GAP was partially supported by JPMorgan Chase I \& Co under J.P. Morgan A.I. Research Awards in 2019 and 2021 and by the EPSRC, grant number EP/P031587/1.
TD is supported by an EPSRC PhD Scholarship.  

}

\appendix

\section{Appendix}\label{sec:Appendix}

The theory of rough paths has proven to be useful in a remarkable variety of fields, stretching from machine learning \cite{chevyrev2016primer}, to regularity theory \cite{hairer2014theory}, then on to variational principles \cite{Leahy2020}, and even further to algebraic topology \cite{Giusti2022} and renormalization theory \cite{Bruned2019}. For a modern introduction to the theory of geometric rough paths, see \cite{Hairer2020}.

The following appendix contains a brief foray into the basic theory of rough paths, following primarily  \cite{Lyons1998,friz2010multidimensional,Hairer2020,Davie2008,lejay2009yet}. It focuses on geometric rough paths, explained from an operational perspective that concentrates on aspects of the theory pertinent to the present paper.\footnote{The discussions in this paper will not distinguish between geometric rough paths and weak geometric rough paths.} 

In particular, the appendix includes the useful theorem \ref{friztheorem} of Friz and Oberhauser \cite{Friz2009}. This theorem is pivotal for the results of the present paper; it describes how the signature of a geometric rough path and its specification as a Lipshitz vector field transform together. Operationally, this theorem formulates in the language of rough paths how additional drift terms arise in the smooth approximations of SDEs discussed in $\cite{Ikeda1981, Wong1965a}$.

The appendix also contains details about coloured noise that are relevant to the present application. Namely, it discusses how coloured noise can be interpreted as a rough path and how the white noise limit can produce modifications to the L\'evy area. The final part of the appendix discusses the numerical methods used in this paper.

\subsection{Key definitions from the theory of rough paths}


In building a robust theory of controlled ordinary differential equations, Lyons \cite{Lyons1998} recognised that solution paths would need to be enhanced by higher order objects based on iterated integrals. These enhanced paths are called \emph{rough paths}. 
A rough path is defined in terms of an ordered series of data known as the \emph{signature} $\mb{X} := (X_t - X_s, \mathbb{X}_{s,t}, \ldots)$ beginning with the vector valued \emph{trace}, $X_t$, and the 2-tensor \emph{enhancement} $\mathbb{X}_{s,t}$, in the series $(X_t - X_s, \mathbb{X}_{s,t}, \ldots)$ of elements of a truncated tensor algebra. The higher terms in the signature refer to aspects of the rough path of lower regularity. In this paper we will focus on two-term signatures which correspond to continuous functions with Hölder exponent between $\frac{1}{3}< \alpha < \frac{1}{2}$. This quadratic class includes almost surely all realisations of Brownian motion and, as a consequence of the Dambis-Dubins-Schwarz theorem,%
\footnote{See Theorem 4.6 \cite{Karatzas1991}, which specifies when continuous local martingales $M_t$ may be written almost surely equal to time changed Brownian motion $B_{\langle M \rangle_t}$. Strong solutions to SDE are necessarily semimartingales and decompose into continuous local martingales plus a finite variation function. } it also includes the solutions of SDE  discussed in Section \ref{sec:3}.

Let $X : [0,T] \rightarrow V$ be a curve in a Banach space $V$ that is $\alpha$-Hölder continuous for $\frac{1}{3}<\alpha < \frac{1}{2}$. Also let there be a map%
\footnote{$\mathbb{X}$ is known as the \emph{enhancement} (or \emph{area process}) of the curve $X_t$.} 
 $\mathbb{X}:\Delta^2_T \rightarrow V \otimes V$
from the $2$-simplex $\Delta^2_T := \{(s,t) : 0 \leq s \leq t \leq T\}$,  that satisfies the following two relations: 
\begin{align}
\begin{split} 
\mathbb{X}_{s, t}-\mathbb{X}_{s, u}-\mathbb{X}_{u, t}=(X_u-X_s) \otimes (X_t-X_u) \qquad \text{(Chen's identity)}\,,
\\ 
\|\mathbb{X}\|_{2 \alpha} := \sup _{s \neq t \in[0, T]} \frac{\left|\mathbb{X}_{s, t}\right|}{|t-s|^{2 \alpha}}<\infty 
\qquad \text{(Hölder continuity)}\,,
\label{6.1}
\end{split}
\end{align} 
(where $X_{s,r}$ is shorthand for the increment $X_r - X_s$) then one says that the pair
\[\mathbf{X} = (X, \mathbb{X}) \in V \oplus [V\otimes V], \] 
is a \emph{rough path} of regularity $\alpha$. The components of the tensor $\mathbb{X}^{ij}$  assign values for the iterated integrals $\int_{s}^{t} X_{s, r}^{i} \dn X_{r}^{j}$. For finite dimensional $V$, any norm induces a norm of tensor products, which enables one to define $|\mathbb{X}|$. Norms on $V \otimes V$ in infinite dimensions require additional constructions. These constructions are not discussed in the present paper, which is restricted to the finite-dimensional case. 

It is crucial to remember that $\mathbb{X}$ defines the iterated integrals, not the other way round, since the fundamental premise is that $X$ is too badly behaved to be used as a direct definition of integration. We enforce Chen's identity in \eqref{6.1} to ensure additivity of limits of integrals, as 

$$\int_t^t X_{t,r}\dn X_r = 0, \quad \int_s^t X_r\dn X_r +  \int_t^u X_r\dn X_r = \int_s^u X_r\dn X_r.$$

In practice, one can expect that the formulation of the problem at hand would motivate the choice of $\mathbb{X}$. Examples of canonical choices for $\mathbb{X}^{ij}$ may be obtained by using Riemann-Stieltjes (for the trivial case of smooth paths), Young (for $\alpha > \frac{1}{2}$ Hölder functions)\cite{Young1936} or a type of stochastic integral. Thus, the canonical choices depend on the smoothness of the path $X$, or on conditions such as adaptedness of a filtration. 

We say that $\mathbf{X}$ is a \emph{geometric rough path} provided in addition to the previous conditions $\mathbb{X}$ satisfies 
\[\operatorname{Sym}(\mathbb{X}_{s,t}) = \frac{1}{2}X_{s,t}\otimes X_{s,t}\,.\] 
This identity is analogous to insisting that integration by parts holds without modifications (as in Stratonovich, rather than It\^o integration). Formally, the sum $\mathbb{X}_{s, t}^{i j}+\mathbb{X}_{s, t}^{j i}$ tells us what to expect for the value of the integrals $\int_{s}^{t} X_{s, r}^{i} \dn X_{r}^{j}+\int_{s}^{t} X_{s, r}^{i} \dn X_{r}^{j}$. An integration by parts identity means that one may manipulate the previous sum of integrals as 
\[
\int_{s}^{t} \dn \left(X^{i} X^{j}\right)_{r}-X_{s}^{i} X_{s, t}^{j}-X_{s}^{j} X_{s, t}^{i} = X^i_{s,t}X^j_{s,t} := (X \otimes X)^{ij}
\,.\]

To ensure that the values of these integrals may be determined by the enhancement, rather than the other way round, we enforce the conditions $\operatorname{Sym}(\mathbb{X}_{s,t}) = \frac{1}{2}X_{s,t}\otimes X_{s,t}$ on $\mathbb{X}$ directly to ensure this integral has the desired property. Enforcing these conditions on $\mathbb{X}$ produces the desired properties of the interpretation of integral.

The space of geometric rough paths forms a Lie subgroup of the truncated tensor algebra $T^{(N)}(V) := \bigoplus_{n=0}^{N}(V)^{\otimes n}$ with group multiplication given by the tensor product. Since $\left[ \frac{1}{\alpha}\right] = 2$ we consider $T^2(\mathbb{R}^d)$. $\mathbf{X}_t = (1, X_t, \mathbb{X}_{0,t})$ is then a Hölder continuous path in the Lie group $G^2(\mathbb{R}^d)$ that satisfies Chen's relations, with $\mathbf{X}_{s,t} := \mathbf{X}_{0,s}^{-1} \otimes \mathbf{X}_{0,t}$. The Lie algebra $\mathcal{G}^2(\mathbb{R}^d)$ is generated by the linear combinations of tensors and their Lie bracket is given by the commutator of tensor products, modulo any triple iterated Lie brackets. The exponential map, its inverse $\log$ and inversion for $G^2(\mathbb{R}^d)$ are defined by truncated power series, as follows,

\begin{equation}\begin{array}{l}
\log (1+X+\mathbb{X}) := X+\mathbb{X}-\frac{1}{2} X \otimes X \,,\\
\\
\quad \exp (X+\mathbb{X}) := 1+X+\mathbb{X}+\frac{1}{2} X \otimes X \,,\\
\\
\quad (1+X+\mathbb{X})^{-1} := 1 - X - \mathbb{X} + X \otimes X \,.
\end{array}\end{equation}

These are global diffeomorphisms since $\mathcal{G}^2(V)$ is nilpotent, where $X + \mathbb{X}$ is shorthand for $(X,\mathbb{X})$. A consequence of the condition for being geometric is that when confining ourselves to this class (such as in Section \ref{sec:2} in order to define rough variational principles) the data required to distinguish types of "noise" when interpreted as a geometric rough path are the trace $X$ and it's L\'evy area $\operatorname{Alt}(\mathbb{X})$. {\color{black}Although not used in this paper, we make a quick note of two remarkable properties that the signature possesses:\\ (i) The signature is invariant under endpoint-preserving time reparametrisations (i.e. pullbacks); and \\(ii) The signature of the time reversed path $X_{-t}$ is the group inverse of the signature in $G^2(\mathbb{R}^d)$.} 

We are now in a position to define the rough integral used in the variational principle \eqref{varprinciple}. For this purpose we define a \emph{controlled rough path}. Given a path $X \in C^\alpha_T(\mathbb{R}^d)$, we say that $Y \in C^\alpha_T(\mathbb{R}^n)$ is controlled by $X$ if there exists some $Y' \in C^\alpha_T(\mathcal{L}(\mathbb{R}^d, \mathbb{R}^n))$ such that the following equation holds,

\begin{equation}Y_{s,t} = Y'_{s}X_{s,t} + R^Y_{(s,t)}\,,\label{gubinelli}\end{equation}

with $\|R^Y_{(s,t)}\|_{2 \alpha} \leq \infty$. We then write that $Y \in D_{X,T}$, where $D_{X,T}$ is the space of paths controlled by $X$. The object $Y'$ in \eqref{gubinelli} is known as the Gubinelli derivative \cite{Gubinelli2004}. In general, the Gubinelli derivative may not be unique. For example, when $Y$ is smooth and we have the identity path $X_{s,t} = t-s$ then a valid choice of Gubinelli derivative for $Y$ would be $0$ and in fact any smooth enough function would suit the purpose. In many situations, the existence of a $Y'$ is of primary importance, rather than its explicit expression. Nevertheless, we shall give well-defined examples of a type of calculus with Gubinelli derivatives for particular classes of rough paths. 

The case that we encounter in Proposition \ref{rdeproposition} is that $Y = F(X)$, with $F$ a differentiable function. It then follows that the ordinary derivative satisfies \eqref{gubinelli}  as a consequence of the first order Taylor expansion. 

For controlled paths $(Y,Y'), (Z,Z')$ the Gubinelli derivative obeys a Leibniz rule for the rough path $W = YZ$ \cite{Hairer2020}[Cor. 7.4], other identities for it can be found in \cite{Armstrong2022}.

Given a controlled rough path $\mb{Y} = (Y, Y') \in D_{X, T}$ we define the rough integral as

\begin{equation}\int_{s}^{t} Y_r \dn \mathbf{X}_r {:=} \lim _{|\mathcal{P}| \rightarrow 0} \sum_{[s, t] \in \mathcal{P}}\left(Y_{s} X_{s, t}+Y_{s}^{\prime} \mathbb{X}_{s, t}\right)\,,\label{roughintegral}\end{equation}

where $\mathcal{P}$ is a suitable partition of the interval. The motivation of this definition is as follows. At leading order in the difference $X_t - X_s$, a Riemann sum is defined in the form: 

$$\int_s^t F(X_r) \dn X_r \approx \sum_{[s, t] \in \mathcal{P}} F(X_s)(X_t - X_s).$$

Given sufficient regularity, this definition would correspond to the Young integral, cf. \cite{lejay2009yet}. This definition corresponds to making the zero order (rectangular) approximation of the integral through $F(X_r) \approx F(X_s)$. If this integral fails to converge due to poor behaviour of $X$ one might try the next order approximation, of the form $F(X_r) \approx F(X_s) + F'(X_s)(X_r - X_s)$. We leave the zeroth order portion of our approximation untouched, but we specifically group the first order term in a rather suggestive way

$$\int_s^t F(X_r) \dn X_r \approx \sum_{[s, t] \in \mathcal{P}} F(X_s)(X_t - X_s) + F'(X_s)X_{s,r}\dn X_r .$$

Here, we use the $X_{s,r}$ shorthand again. One now recognises the prescence of $X_{s,r}\dn X_r$ as what we called $\mathbb{X}_{s,t}$, and we abstract $F'(X_s)$ into the Gubinelli derivative of $Y$ when dealing with arbitrary integrands, concluding in definition \eqref{roughintegral}. The moral is that the rough integral is a form of the Riemann sum limit taken to higher order. 

Rough integrals themselves are controlled by $X$ with Gubinelli derivative equal to their integrand. This is a consequence of definitions \eqref{roughintegral}, \eqref{gubinelli} and the fact that $Y' \mathbb{X}$ is of order $o(|t-s|^{2\alpha})$.

We say that $Y$ is a solution of the rough differential equation
\begin{equation}\dn Y = f(Y) \dn t + g(Y) \dn \mb{X}, \quad Y_0 = \xi,\label{RDE}\end{equation}
if and only if $\mb{Y} = (Y, g(Y)) \in D_{X, T}$ and the following integral equation is satisfied in the sense of \eqref{roughintegral}

\begin{equation}Y_t = \xi + \int_0^t f(Y_r)\dn r + \int_0^t g(Y_r)\dn \mb{X}_r\,.\end{equation}

We also introduce the following equivalent characterisation of a solution, known as the Davie formulation \cite{Davie2008}. The Davie formulation brings forward the explicit dependence of $\mathbb{X}$ which is particularly helpful when one wishes to separate terms that contribute perturbations of the L\'evy area in Proposition \ref{rdeproposition}.

\begin{equation}\delta Y_{st} - \int_{0}^tf(Y_r)\dn r - g(Y_s)\delta X_{st} - Dg(Y_s)g(Y_s)\mathbb{X}_{st} = o(|t-s|).\end{equation}

Since rough differential equations are interpreted in the integral sense, it follows that if $Y$ solves \eqref{RDE} then it admits a Gubinelli derivative $g(Y)$. In an intuitive sense, one may say that the Gubinelli derivative "differentiates with respect to the noise".

\subsection{Antisymmetric additions to the signature tensor}

The following theorem found in \cite{Friz2009} tells us that antisymmetric perturbations to the signature tensor of a rough path will appear as correction terms in a RDE driven by paths possessing a non-canonical L\'evy area.

\begin{theorem}[Friz, Oberhauser \cite{Friz2009}] \label{friztheorem}
Let $\boldsymbol{X} :\Delta^2_T \rightarrow \mathcal{G}^{[\frac{1}{\alpha}]}(\mathbb{R}^d)$ be a geometric $\alpha$-H{\"o}lder rough path. Fix $\mathsf{s} \in \mathcal{G}^{[\frac{1}{\alpha}]}(\mathbb{R}^d) \cap \mathbb{R}^d \otimes \mathbb{R}^d$ and define $\widetilde{\boldsymbol{X}}_{s,t} := \boldsymbol{X}_{s,t} + \mathsf{s}(t-s)$. \newline

Let $V = V_i$, $1 \leq i \leq d$ be a collection of differentiable vector fields with Lipschitz derivatives on $\mathbb{R}^e$. The  solution of the RDE

\begin{equation}\dd x = V_0\dd t + V(x)\dd \widetilde{\boldsymbol{X}},\end{equation}

coincides with the solution of the RDE

\begin{equation}\dd x = V_0\dd t + \frac{1}{2}\sum_{1 \leq i \leq j \leq d}\mathsf{s}^{ij}[V_i,V_j]\dd t + V(x)\dd \boldsymbol{X}.\label{eq:a9}\end{equation}

Where $[V_i,V_j]$ denotes the commutator of vector fields\footnote{We alert the reader to the difference in summing convention between \eqref{eq:a9} and \cite{Friz2009, Caruana2009}, hence the additional factor of $\frac{1}{2}$ that must be included here.}. \end{theorem}

In Section \ref{sec:2}, Theorem \ref{friztheorem} is used to derive the Lie-Poisson equations. One can choose to apply Theorem \ref{friztheorem} to the rough Hamilton equations \eqref{roughsymplectic} and obtain \eqref{hamiltonlevy} as discussed in Section \ref{2.3}. Alternatively, one can use Theorem \ref{friztheorem} directly on the Lie-Poisson equation \eqref{eptilde}. 
In this case, one defines the action of the vector fields $V_i \in \mathfrak{X}$ on the variable $\mu\in \mathfrak{X}^*$ as $V_i(\mu) = \operatorname{ad}^*_{\xi_i}\mu$. Direct calculation then reveals that 
\begin{equation}
[V_i, V_j](\mu) := V_i(V_j(\mu)) - V_j(V_i(\mu)) = 
\ad^*_{\xi_i}(\ad^*_{\xi_j}\mu) - \ad^*_{\xi_j} (\ad^*_{\xi_i} \mu) = -\ad^*_{[\xi_i, \xi_j]}\mu
\,. \label{lpapplication}\end{equation}


The rough Hamilton-Pontryagin constraints in \eqref{velocity} and \eqref{othervelocity} in variational principles \eqref{varprinciple} and \eqref{othervarprinciple} can be seen as equivalent by applying Theorem \ref{friztheorem} to the rough velocity equation $\dn x = u\dn t + \xi \dn \widetilde{\mb{Z}}$. Now Theorem \ref{friztheorem} applies for finite dimensional RDE only. However, generalisations to rough partial differential equations (RPDE) exist, see \cite{Caruana2009}, although this generalisation is not an issue for the purposes of this paper. The equivalence of the vector fields \eqref{velocity} and \eqref{othervelocity} remains true for infinite dimensional Lie groups, as well. Consequently, it is also applicable for fluid dynamics. For the fluid dynamics formulations, see \cite{Leahy2020,coghi2021robust}.

\subsection{Example: Coloured noise as a rough path}\label{sec:colouredappendix}

In this section we examine the approximation of white noise by coloured noise as an example of the Wong-Zakai correction terms of the type in \eqref{ikedawatanabe}. Although this approximation problem arising in stochastic analysis can be discussed without invoking rough paths theory, lifting it into the rough path context allows us to understand how its failure to converge to the signature of Stratonovich Brownian motion introduces L\'evy area correction terms to the Wong-Zakai convergence theorem.

Let $y_t \in \mathbb{R}^n$ be the solution of the SDE

\begin{equation}\dd y_t = -\frac{1}{\varepsilon^2}Ay_t\dd t + \frac{1}{\varepsilon}D \dd B_t\label{SDE}.\end{equation}

Where $A$ and $D$ are constant $n \times n$ matrices whose eigenvalues have positive real part. The distinction between It\^o vs Stratonovich in this case is unnecessary due to the noise being additive, but it will be useful to calculate with the It\^o integral and transfer back to Stratonovich to state results. We define the process $B^\varepsilon_t := \int_0^t\frac{y_s}{\varepsilon}\dd s$ and lift this to a rough path $\mathbf{X} = (B^\varepsilon, \mathbb{B}^\varepsilon)$ with the canonical choice of its iterated integrals in the Young sense (this is well defined, since $B^\varepsilon$ has finite variation).

\noindent {\textbf{Claim:}} The trace $B^\varepsilon_t$ converges to $A^{-1}DB_t$.

{\textbf{Proof:}} In the SDE \eqref{SDE}, $y_t$ is an Ornstein-Uhlenbeck process. Hence, it be rearranged as   

$$\frac{y_t}{\varepsilon}\dn t= -\varepsilon A^{-1}dy_t + A^{-1}D  \dn B_t.$$ 

In the integral sense, the SDE \eqref{SDE} is equivalent to:

$$\int_0^t \frac{y_t}{\varepsilon}\dn t= \int_0^t -\varepsilon A^{-1}dy_t + \int_0^t A^{-1}D  \dn B_t.$$

To prove convergence, it suffices to show that the term $\int_0^t -\varepsilon A^{-1}dy_t$ vanishes as $\varepsilon \rightarrow 0$ in a suitable topology. For this purpose, the following Borel type inequality is available \cite{Adler1990,Pavliotis2005,Pavliotis2003},

\begin{equation}
\mathbb{E}\left(\sup _{0 \leq t \leq T}|y_t|^{2 n}\right) \leq\left(\frac{k_D}{k_A}\right)^{n} C_{n}\left[1+\left(\log \left(\frac{k_AT}{\epsilon^{2}}\right)\right)^{n}\right].
\label{Greg's ID}
\end{equation}

Here, $k_A, k_D$ are constants depending on the spectral radius of the matrices $A,D$. Now one finds,
$$\mathbb{E} \left[\sup_{0 \leq t \leq T}\left|\int_0^t -\varepsilon A^{-1}dy_t\right|^2\right] = \varepsilon^2 \mathbb{E} \left[\sup_{0 \leq t \leq T}\left|A^{-1}y_t \right|^2\right] \leq \varepsilon^2 C\mathbb{E} \left[\sup_{0 \leq t \leq T}\left|y_t \right|^2\right] \leq \varepsilon M \left( 1 + \log\left(\frac{kT}{\varepsilon^2}\right) \right) \rightarrow 0.$$

Thus the integral of $\frac{y_t}{\varepsilon}$ converges uniformly in mean square to a linear transformation of Brownian motion.

It remains to prove convergence of the area process $\mathbb{B}^\varepsilon$. For the proof, we shall use the Ergodic Theorem to show $\mathbb{B}^\varepsilon \rightarrow (A^{-1}D)^{\otimes 2}(\mathbb{B}^{\text{Strat}}) + \mathsf{s}(t-s)$ and shall calculate $\mathsf{s} = \Sigma (A^{-1})^T - A^{-1}\Sigma$, where $\Sigma$ is the symmetric covariance matrix of the stationary Gaussian distribution for $y_{\varepsilon^{-2}t}$. The notation $(A^{-1}D)^{\otimes 2}(\mathbb{B}^{\text{Strat}})$ denotes the canonical area process of Stratonovich Brownian motion defined as $\mathbb{B}^{\text{Strat}}_{ij} := \int B^i \circ \dn B^j$ multiplied by the tensor product of linear maps $A^{-1}D \otimes A^{-1}D$.

{\textbf{Proof:}} We follow the techniques of similar calculation appearing in \cite{Friz_2015, Hairer2020}. By definition we know that 
\[\dn B^\varepsilon_t = \frac{y}{\varepsilon}\dn t
\quad\hbox{and}\quad
\dn y_t =  -\frac{1}{\varepsilon^2}Ay\dn t+ D\dn B_t
\,.\]
Hence, we can rearrange the differentials to obtain 
\[
\dn B_t^\varepsilon = -\varepsilon A^{-1}\dn y_t + A^{-1}D\dn B_t
\,.\]
This expression is well defined when considering the respective integral equations. By using the identity \eqref{Greg's ID} above, we can thus write:

$$\mathbb{B}^\varepsilon_{t} := \int_0^tB^\varepsilon_s \otimes \dn B^\varepsilon_s = \int_0^tB^\varepsilon_s \otimes A^{-1}D\dn B_s - \varepsilon\int_0^tB^\varepsilon_s \otimes A^{-1} \dn y_s.$$

Next, integration by parts and the differential relation $\dn B^\varepsilon_t = \frac{y}{\varepsilon}dt$ used for defining $B_t^\varepsilon$ together yield

$$\mathbb{B}^\varepsilon_{t} = \int_0^tB^\varepsilon_s \otimes A^{-1}D\dn B_s + B^\varepsilon_t \otimes \left(-\varepsilon A^{-1} y_t\right) + \int_0^ty_s \otimes A^{-1}y_s\dn s.$$

By the dominated convergence theorem, the first term converges precisely to 
\[
(A^{-1}D)^{\otimes 2}(\mathbb{B}^{\text{It\^o}}) = (A^{-1}D)^{\otimes 2}(\mathbb{B}^{\text{Strat}}) - (A^{-1}D)^{\otimes 2}\Big(\frac{t-s}{2}I\Big)
\,.\]
The second term converges to zero, by the uniform $L^2$ bound obtained on $-\varepsilon A^{-1} y_t$ and the fact that $B_t^\varepsilon$ is bounded, since it is a convergent sequence. We will now show the final integral converges by using the Ergodic Theorem, as in the statement in \cite{Kallenberg2002}. \newline

For a given $\varepsilon$, define a new Brownian motion $\tilde{B}_t = \varepsilon B_{\varepsilon^{-2}t}$ and consider the process $\tilde{y}$ defined as the solution of the SDE $\dn \tilde{y} = -A\tilde{y}\dd t + D\dn \tilde{B}$. 

It can be shown that $y_t$ and $\tilde{y}_{\varepsilon^{-2}t}$ are pathwise equal for the same initial conditions. Furthermore, $\tilde{y}$ is an Ergodic Markov process and the stationary solution $Y_t$ is a zero mean Gaussian with covariance matrix given by 

$$\Sigma = \int_{-\infty}^0\exp(\rho A)DD^T\exp(\rho A^T)\dn \rho \,.$$

We denote the law of $Y$ by $\nu$ and write

$$\int_0^tf(y_s)\dn s = \int_0^tf(\tilde{y}_{\varepsilon^{-2}s})\dn s = \varepsilon^2\int_0^\frac{t}{\varepsilon^2}f(\tilde{y}_{\tau})\dn \tau = t\frac{1}{T}\int_0^Tf(\tilde{y}_{\tau})\dn \tau \rightarrow t \int f\dn \nu .$$

For our purposes we take $f(y_s) = y_s \otimes A^{-1}y_s = I \otimes A^{-1}(y_s \otimes y_s)$, the integral $\int f\dn \nu$ is the expectation 
\[
\mathbb{E}[f(Y)] = \mathbb{E}[I \otimes A^{-1}(Y \otimes Y)] = I \otimes A^{-1}(\mathbb{E}[Y \otimes Y]) = I \otimes A^{-1}(\Sigma) = \Sigma (A^{-1})^T\,.
\]
Here we have made use of the matrix identity $M \otimes N \left(X\right) = MXN^T$ for $M \in \mathbb{R}^{n \times n}, N \in \mathbb{R}^{m \times m}, X \in \mathbb{R}^{n}\otimes \mathbb{R}^m$. \newline

We have now shown $\mathbb{B}^\varepsilon_{s,t} \rightarrow (A^{-1}D)^{\otimes 2}(\mathbb{B}^{\text{Strat}}_{s,t}) + (t-s)\left(\Sigma (A^{-1})^T - \frac{1}{2}A^{-1}DD^T(A^{-1})^T\right)$. Since the limit of a geometric rough path remains geometric, the symmetric parts must converge to the tensor product of the traces. That is,
the symmetric parts converge as

$$\operatorname{Sym}(\mathbb{B}_t^\varepsilon) = \frac{1}{2}B_t^\varepsilon \otimes B_t^\varepsilon \rightarrow (A^{-1}D)^{\otimes 2} \Big(\frac{1}{2}B_t \otimes B_t\Big) .$$

Convergence of the symmetric parts implies that the matrix term:  
\[
\Sigma (A^{-1})^T - \frac{1}{2}A^{-1}DD^T(A^{-1})^T
\,.\]

has symmetric part zero and the correction to the L\'evy stochastic area can only be antisymmetric. $\mathsf{s}^T = -\,\mathsf{s}$ allows us to write the more compact form, \[\mathsf{s} = \Sigma (A^{-1})^T - A^{-1}\Sigma\,.\]

\subsection{Numerical methods}\label{sec:methods}
Here we descibe in more detail the numerical scheme used in obtaining the results of Section \ref{sec:4}. We solve equation \eqref{srblevy} using an Implicit Midpoint Munthe-Kaas method. The method described below is an adaptation of the highly versatile numerical scheme in \cite{Luesink2021} to integrate stochastic Lie-Poisson equations in a Casimir preserving manner. 

A key feature of the double-bracket dissipative form of the Lie-Poisson equations used here is that their dynamics is coadjoint motion. That is, their solution can expressed as $\mu_t = \operatorname{Ad}^*_{g(t)}\mu_0$ for some curve $g(t)$ in $G$. The local representation of $g(t) = \exp(\sigma(t))$ for $\sigma \in \mathfrak{g}$ and the identity $\operatorname{Ad}_{\exp} = \exp(\ad)$ allows us to find a solution by solving the equation 
\[
\frac{d\sigma}{d t} =\frac{\operatorname{ad}_{\sigma}}{1-\exp \left(-\operatorname{ad}_{\sigma}\right)}\left( \frac{\delta }{\delta \mu}\Big(h {\color{black}-} \frac{1}{2}\mathsf{s}^{ij}\{\Phi_i,\Phi_j\} + \vartheta \ad^*_{\frac{\delta h}{\delta \mu}}\mu\Big)\right)
\,.\]
The operator $\frac{\operatorname{ad}_{\sigma}}{1-\exp \left(-\operatorname{ad}_{\sigma}\right)}$ can be approximated by the identity operator introducing errors of order higher than one \cite{Luesink2021}[Sec. 4].  

It is shown in \cite{Luesink2021} that consistency of this numerical method requires 

\begin{equation}
    \sigma_{n}= \left(  \frac{\delta h}{\delta  \mu}\left(\frac{\mu_{n}+\mu_{n+1}}{2}\right) - \frac{1}{2}\mathsf{s}^{ij}[\xi_i, \xi_j] + \vartheta \frac{\delta }{\delta \mu}\left(\ad^*_{\frac{\delta h}{\delta \mu}}  \mu \right)\left(\frac{\mu_n+\mu_{n+1}}{2}\right)\right)\Delta t + \sum_i \Delta W^i \xi_i .
\label{energyconserve}\end{equation}

We solve for each $\sigma_n$ in \eqref{energyconserve} using the chord method, and then iterate $\mu_{n+1} = \operatorname{Ad}^*_{\exp(\sigma_n)}\mu_n$.

The coloured noise case, \eqref{srbcoloured}, is solved with the same discretisation. The Browian increment $\Delta W$ is an ordinary time step $\Delta t$ times $\frac{1}{\varepsilon}(y_{n+1} - y_n)$. We generate the coloured noise process $y_t$ by solving \eqref{coloured} with a two-step implicit order algorithm using the Python package called  \emph{sdeint}.

For details of its derivation and the many useful properties of this scheme, we refer the reader to \cite{Luesink2021}. \newpage

\subsection{Additional Figures}\label{sec:figs}

Here we provide another viewpoint of bifurcations demonstrated in \ref{surf2} as well as illustrate the complex behaviour arising when the L\'evy area has components in several directions.

\begin{figure}[H]
    \centering
    \includegraphics[width=0.4\linewidth]{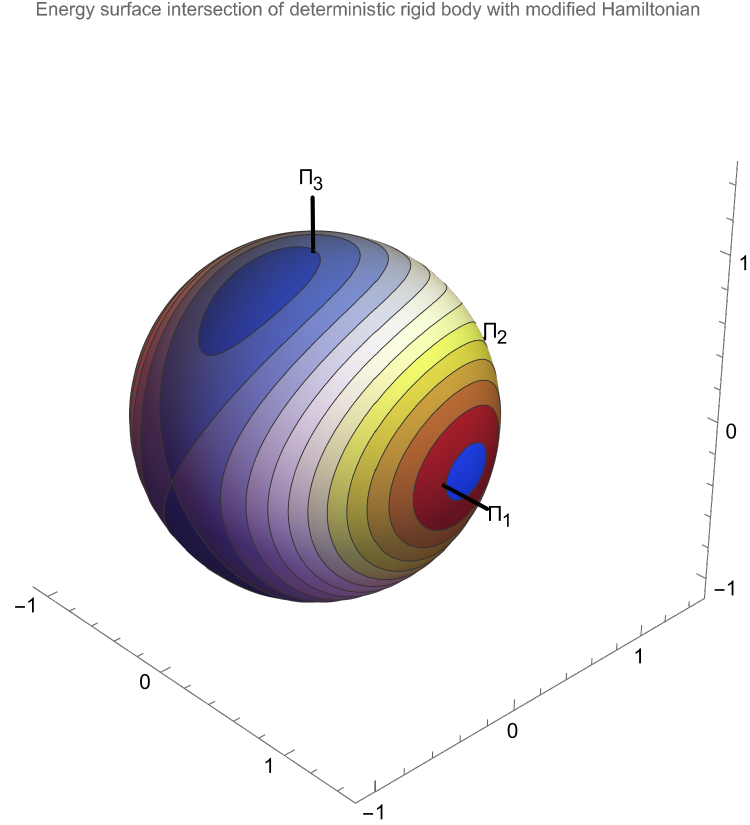}
    \caption{From this viewpoint, one sees that the figure of eight pair of homoclinic orbits that forms on the opposite side of the angular momentum sphere to that in Figure \ref{surf2} is rotated by $\pi/2$.} 
\label{surf3}\end{figure}

\begin{figure}[H]
  \centering
  \begin{minipage}[b]{0.4\textwidth}
    \includegraphics[width=\textwidth]{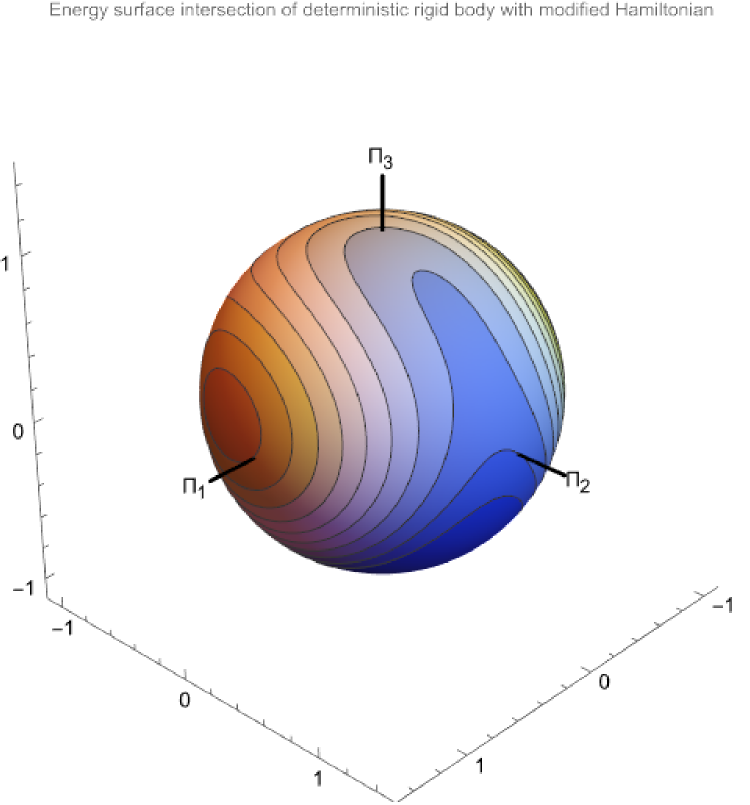}
    \caption{ }
  \label{surf4}\end{minipage}
  \hfill
  \begin{minipage}[b]{0.4\textwidth}
    \includegraphics[width=\textwidth]{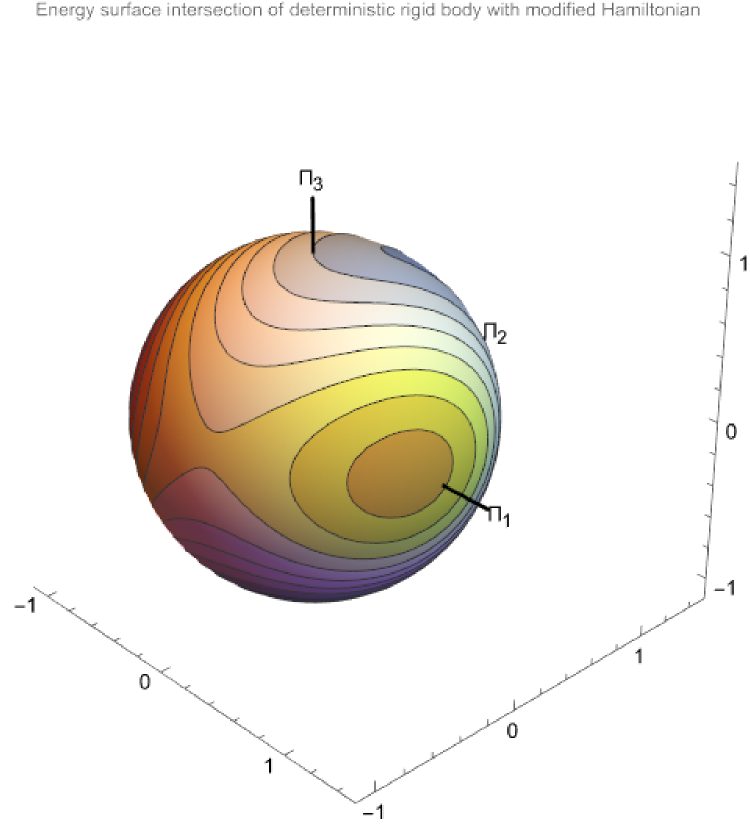}
    \caption{ }
  \label{surf5}\end{minipage}
  \caption*{Two perspectives of intersections of Hamiltonian $\widetilde{h}(\Pi) = \frac{1}{2}\Pi \cdot \mathbb{I}^{-1} \Pi + 0.0625\Pi \cdot \mathbf{\hat{e}}_1 - 0.0625\Pi \cdot \mathbf{\hat{e}}_2  - 0.125\Pi \cdot \mathbf{\hat{e}}_3$ with the angular momentum sphere $|\Pi|^2=const$ are shown. These intersections are interpreted as the effect of the anisotropic L\'evy area correction for $\sigma = 0.05$ and $\mathsf{s}^{12} = -50, \mathsf{s}^{13} = 25, \mathsf{s}^{23} = 25$ on the drift dynamics. Locations of all fixed points have been shifted and the phase portrait has no discrete symmetry.}
\end{figure}


\bibliographystyle{plainurl}
\bibliography{bibliography.bib}
\end{document}